\documentclass{pasj01}
\usepackage{lscape} 

\begin{document}
\SetRunningHead{Author(s) in page-head}{Running Head}
%\Received{}%{yyyy/mm/dd}
%\Accepted{}%{yyyy/mm/dd}
%\Published{}%{yyyy/mm/dd}

\title{Physical properties of near-Earth asteroids with a low delta-{\it v}: Survey of target candidates for the Hayabusa2 mission}

%%% begin:list of authors
\author{%
   Sunao \textsc{Hasegawa},\altaffilmark{1,*}
   Daisuke \textsc{Kuroda},\altaffilmark{2}
   Kohei \textsc{Kitazato},\altaffilmark{3}
   Toshihiro \textsc{Kasuga},\altaffilmark{4,5}
   Tomohiko \textsc{Sekiguchi},\altaffilmark{6}
   Naruhisa \textsc{Takato},\altaffilmark{7}
   Kentaro \textsc{Aoki},\altaffilmark{7}
   Akira \textsc{Arai},\altaffilmark{8}
   Young-Jun \textsc{Choi},\altaffilmark{9}
   Tetsuharu \textsc{Fuse},\altaffilmark{10}
   Hidekazu \textsc{Hanayama},\altaffilmark{11}
   Takashi \textsc{Hattori},\altaffilmark{7}
   Hsiang-Yao \textsc{Hsiao},\altaffilmark{12}
   Nobunari \textsc{Kashikawa},\altaffilmark{13}
   Nobuyuki \textsc{Kawai},\altaffilmark{14}
   Kyoko \textsc{Kawakami},\altaffilmark{1,15}
   Daisuke \textsc{Kinoshita},\altaffilmark{12}
   Steve \textsc{Larson},\altaffilmark{16}
   Chi-Sheng \textsc{Lin},\altaffilmark{12}
   Seidai \textsc{Miyasaka},\altaffilmark{17}
   Naoya \textsc{Miura},\altaffilmark{18}
   Shogo \textsc{Nagayama},\altaffilmark{4}
   Yu \textsc{Nagumo},\altaffilmark{5}
   Setsuko \textsc{Nishihara},\altaffilmark{1,15}
   Yohei \textsc{Ohba},\altaffilmark{1,15}
   Kouji \textsc{Ohta},\altaffilmark{19}
   Youichi \textsc{Ohyama},\altaffilmark{20}
   Shin-ichiro \textsc{Okumura},\altaffilmark{21}
   Yuki \textsc{Sarugaku},\altaffilmark{8}
   Yasuhiro \textsc{Shimizu},\altaffilmark{22}
   Yuhei \textsc{Takagi},\altaffilmark{7}
   Jun \textsc{Takahashi},\altaffilmark{23}
   Hiroyuki \textsc{Toda},\altaffilmark{2}
   Seitaro \textsc{Urakawa},\altaffilmark{21}
   Fumihiko \textsc{Usui},\altaffilmark{24}
   Makoto \textsc{Watanabe},\altaffilmark{25}
   Paul \textsc{Weissman},\altaffilmark{26}
   Kenshi \textsc{Yanagisawa},\altaffilmark{27}
   Hongu \textsc{Yang},\altaffilmark{9}
   Michitoshi \textsc{Yoshida},\altaffilmark{7}
   Makoto \textsc{Yoshikawa},\altaffilmark{1}
   Masateru \textsc{Ishiguro},\altaffilmark{28}
   and
   Masanao \textsc{Abe}\altaffilmark{1}
}
 \altaffiltext{1}{Institute of Space and Astronautical Science, Japan Aerospace Exploration Agency, 3-1-1 Yoshinodai, Chuo-ku, Sagamihara 252-5210, Japan}
 \email{hasehase@isas.jaxa.jp}
 \altaffiltext{2}{Okayama Astronomical Observatory, Kyoto University, 3037-5 Honjo, Kamogata-cho, Asakuchi, Okayama 719-0232, Japan}
 \altaffiltext{3}{Research Center for Advanced Information Science and Technology, The University of Aizu, Tsuruga, Ikki-machi, Aizu-Wakamatsu, Fukushima 965-8580, Japan}
 \altaffiltext{4}{Public Relations Center, National Astronomical Observatory of Japan, 2-21-1 Osawa, Mitaka-shi, Tokyo 181-8588, Japan}
 \altaffiltext{5}{Department of Physics, Kyoto Sangyo University, Motoyama, Kamigamo, Kita-ku, Kyoto 603-8555, Japan}
 \altaffiltext{6}{Department of Education, Hokkaido University of Education, 9 Hokumon, Asahikawa-shi, Hokkaido 070-8621, Japan}
 \altaffiltext{7}{Subaru Telescope, National Astronomical Observatory of Japan, 650 North A'õhoku Place, Hilo, HI 9672, USA}
 \altaffiltext{8}{Koyama Astronomical Observatory, Kyoto Sangyo University, Motoyama, Kamigamo, Kita-ku, Kyoto 603-8555, Japan}
 \altaffiltext{9}{Space Science Division, Korea Astronomy and Space Science Institute, 776, Daedeokdae-ro, Yuseong-gu, Daejeon, Korea}
 \altaffiltext{10}{Strategic Planning Department, National Institute of Information and Communications Technology, 4-2-1 Nukuikitama-chi, Koganei, Tokyo 184-8795, Japan}
 \altaffiltext{11}{Ishigakijima Astronomical Observatory, National Astronomical Observatory of Japan, 1024-1 Arakawa, Ishigaki, Okinawa 907-0024, Japan}
 \altaffiltext{12}{Institute of Astronomy, National Central University, 300 Zhongda Rd., Zhongli, Taoyuan City 320, Taiwan, R.O.C.}
 \altaffiltext{13}{Department of Astronomy, The University of Tokyo, 7-3-1 Hongo, Bunkyo-ku, Tokyo 113-0033, Japan}
 \altaffiltext{14}{Department of Physics, Tokyo Institute of Technology, 2-12-1 Ookayama, Meguro-ku, Tokyo 152-8551, Japan}
 \altaffiltext{15}{Department of Earth and Planetary Science, The University of Tokyo, 7-3-1 Hongo, Bunkyo-ku, Tokyo 113-0033, Japan}
 \altaffiltext{16}{Lunar and Planetary Laboratory, University of Arizona, 1629 E University Blvd, Tucson, Arizona 85721-0092, USA}
 \altaffiltext{17}{Tokyo Metropolitan Government, 2-8-1 Nishishinjyuku, Shinjyuku-ku, Tokyo 163-8001, Japan}
 \altaffiltext{18}{Department of Multi-Disciplinary Sciences, The University of Tokyo, 3-8-1 Komaba, Meguro-ku, Tokyo 153-8902, Japan}
 \altaffiltext{19}{Department of Astronomy, Kyoto University, Kitashirakawaoiwake-cho, Sakyo, Kyoto 606-8502, Japan}
 \altaffiltext{20}{Academia Sinica, Institute of Astronomy and Astrophysics, 11F of Astronomy-Mathematics Building, AS/NTU, No.1, Section 4, Roosevelt Rd., Taipei 10617, Taiwan, R.O.C.}
 \altaffiltext{21}{Japan Spaceguard Association, Bisei Spaceguard Center 1716-3 Okura, Bisei, Ibara, Okayama 714-1411, Japan}
 \altaffiltext{22}{Subaru Telescope Okayama Branch Office, National Astronomical Observatory of Japan, 3037-5 Honjo, Kamogata-cho, Asakuchi, Okayama 719-0232, Japan}
 \altaffiltext{23}{Nishi-Harima Astronomical Observatory, University of Hyogo, 407-2, Nishigaichi, Sayo-cho, Sayo, Hyogo 679-5313, Japan}
 \altaffiltext{24}{Center for Planetary Science, Kobe University, 7-1-48, Minatojima-minamimachi, Chuo-Ku, Kobe 650-0047, Japan}
 \altaffiltext{25}{Department of Applied Physics, Okayama University of Science, 1-1 Ridai, Kita-ku, Okayama, Okayama 700-0005, Japan}
 \altaffiltext{26}{Planetary Science Institute, 1700 East Fort Lowell, Suite 106, Tucson, Arizona 85719, USA}
 \altaffiltext{27}{Division of Optical and Infrared Astronomy, National Astronomical Observatory of Japan, 2-21-1 Osawa, Mitaka-shi, Tokyo 181-8588, Japan}
 \altaffiltext{28}{Department of Physical and Astronomy, Seoul National University, Gwanak-ro, Gwanak-gu, Seoul 08826, Korea}
%%% end:list of authors

%% `\KeyWords{}' always has to be placed before `\maketitle'.
 %Do NOT move this preamble from here!
\maketitle

\begin{abstract}
Sample return from the near-Earth asteroid known as 25143 Itokawa was conducted as part of the Hayabusa mission, with a large number of scientific findings being derived from the returned samples.
Following the Hayabusa mission, Hayabusa2 was planned, targeting sample return from a primitive asteroid.
The primary target body of Hayabusa2 was asteroid 162173 Ryugu; however, it was also necessary to gather physical information for backup target selection.
Therefore, we examined five asteroids spectroscopically, 43 asteroids spectrophotometrically, and 41 asteroids through periodic analysis.
Hence, the physical properties of 74 near-Earth asteroids were obtained, which helped the Hayabusa2 backup target search and, also, furthered understanding of the physical properties of individual asteroids and their origins.
\end{abstract}

\section{Introduction}
Cosmochemical, mineralogical, and petrological analyses of extraterrestrial materials such as Apollo and Luna samples, meteorites, and interplanetary dust particles have supplied important clues on their chemical composition, the temperature and pressure conditions they have encountered, and on solar system formation, composition, and evolution.
Therefore, sample return missions from major or minor planets can provide considerable scientific information compared to flyby and/or rendezvous missions.

In the 1990s, the Mu Space Engineering Spacecraft (MUSES)-C mission was implemented with the aim of successful sample return from an asteroid (\authorcite{Fujiwara2000} \yearcite{Fujiwara2000}; \yearcite{Fujiwara2004}).
The MUSES-C spacecraft was launched in 2003, but renamed ``Hayabusa'' after launch.
In 2005, Hayabusa rendezvoused with the S-complex asteroid known as 25143 Itokawa and performed sample collection of material on the asteroid surface \citep{Fujiwara2006}.
Samples taken from Itokawa were brought to Earth in 2010 and analysis was performed.
Those samples provided important planetary science information regarding the meteorite source, the thermal and impact history of Itokawa, the space weathering process on asteroids (\cite{Ebihara2011}; \cite{Nagao2011}; \cite{Nakamura2011}; \cite{Noguchi2011}; \cite{Tsuchiyama2011}; \cite{Yurimoto2011}).

Following the success of the Itokawa rendezvous performed by the Hayabusa spacecraft, a subsequent asteroid exploration project, ``Hayabusa2'', was planned \citep{Tsuda2013}.
The objective of the Hayabusa mission was demonstration of sample return technology; thus, the target body was an object reachable by the Mu-5 launch vehicle.
The purpose of Hayabusa2, which launched in 2014 and is expected to return to Earth in 2020, is sample return from a chondritic asteroid such as a C- or D-complex body.
S-complex asteroids, i.e., of the same spectral type as Itokawa, are excluded.
Thus, the C-complex asteroid 162173 Ryugu was selected as the primary target of Hayabusa2.
Ryugu was chosen because it is a C-complex asteroid having one of the lowest delta-$v$ among the near-Earth asteroids (NEAs) known at the time (delta-$v$ is the velocity adjustment needed to transfer a spacecraft from low-Earth orbit to connect with the asteroid \citep{Shoemaker1978}).
Hayabusa2 rendezvoused with Ryugu in 2018.

Before exploration, the physical properties of the asteroid, such as the rotational period, spin vector, geometric albedo, and spectral type, must be known; hence, the spacecraft design and operation sequence for rendezvous with and sampling from the asteroid can be determined.
Therefore, we performed observations to obtain the physical properties of the prospective candidates for Hayabusa (\cite{Abe2000}; \cite{Ishibashi2000a}; \cite{Ishibashi2000b}; \cite{Ishiguro2003}; \cite{Kaasalainen2003};  \cite{Ohba2003};  \cite{Sekiguchi2003}; \cite{Cellino2005}; \cite{Lederer2005} ; \cite{Mueller2005}; \cite{Mueller2007}; \cite{Nishihara2018}) and Hayabusa2 (\cite{Hasegawa2008}; \cite{Mueller2011}; \cite{Urakawa2011}; \cite{Kim2013}; \cite{Ishiguro2014}; \cite{Kuroda2014}; \cite{Mueller2017}; \cite{Perna2017a}).

Note that the target body of the Hayabusa mission was changed twice, with corresponding launch postponement.\footnotemark[1]
In addition, while it is necessary to know the physical characteristics of the primary target, it is also important to obtain those of backup candidates with consideration of potential mission problems.
Therefore, we conducted observations to acquire the physical properties of backup target candidates for the Hayabusa2 mission.

\footnotetext[1]{The initial target was the E-type asteroid 4660 Nereus; the second target was the E-type asteroid 10302 1989 ML.}

In this paper, we report on the physical properties of the NEAs with low delta-$v$ examined for the Hayabusa2 mission.
The subsequent sections of this paper describe the observations and data reduction procedures (section 2), present the results on the asteroid physical properties (section 3), and discuss the implications of our findings (section 4).

\section{Observations and data reduction procedures}
The asteroid observations were performed from December 2000 to January 2015, using 10 different telescopes at eight sites in Japan and the USA, with photometry and spectroscopy observations.
Five NEA spectra were recorded using the Subaru Telescope on Mt. Mauna Kea, Hawaii, USA (Minor Planet Center (MPC) code: 568) and at the Okayama Astrophysical Observatory in Okayama, Japan (MPC code: 371).
Colors were recorded for 38 NEAs, along with five other asteroids, at the Kiso Observatory in Nagano, Japan (MPC code: 381); the Lulin Observatory in Nantou, Taiwan (MPC code: D35); the UH88 Telescope on Mt. Mauna Kea (MPC code: 568); the Steward Observatory on Mt. Bigelow, Arizona, USA (MPC code: 698); and the Nayoro Observatory in Hokkaido, Japan (MPC code: Q33).
The lightcurves of 39 NEAs and two other asteroids were obtained at the Kiso Observatory; the Lulin Observatory; the Nayoro Observatory; the Ishigakijima Astronomical Observatory at Okinawa, Japan (MPC code: D44); the Nishiharima Astronomical Observatory at Hyogo, Japan (no MPC code; 134$\timeform{D}$20'09''E, 35$\timeform{D}$01'33''N; 450 m); and the Bisei Spaceguard Center at Okayama, Japan (MPC code: 300).

\subsection{Spectroscopic observations}
Asteroid spectroscopic data were recorded by two different instruments mounted on different telescopes: the Faint Object Camera and Spectrograph (FOCAS) \citep{Kashikawa2002} installed at the f/12 Cassegrain focus of the 8.2-m Subaru Telescope, and the Kyoto Okayama Optical Low-dispersion Spectrograph (KOOLS) (\cite{Ohtani1998}; \cite{Ishigaki2004}) attached to the f/18 Cassegrain focus of the 1.88-m telescope at the Okayama Astrophysical Observatory.
Three and two spectrocopic observations of the NEAs were performed using FOCAS and KOOLS, respectively.
The nightly observation details of the spectroscopy are listed in table \ref{tab:spectroscopic circumstances}.

%%%%%%%%%%%%%%%%%%%%%%%%%%%%%%%%%%%%%%%
\begin{longtable}{llllllll}
  \caption{Spectroscopic circumstances of the asteroids.}\label{tab:spectroscopic circumstances}
  \hline
    No. & Name & Date      & $R_{\rm h}$\footnotemark[$*$] & $\Delta$\footnotemark[$*$] & $\alpha$\footnotemark[$*$] & Instrument/ &  Standard star\\
         &     & [YY.MM.DD]&[au]         & [au]     & [$\timeform{D}$] &Telescope    & \\
\endfirsthead
  \hline
    No. & Name & Date & $R_{\rm h}$ & $\Delta$ & $\alpha$ & Instrument & Standard star\\
  \hline
\endhead
  \hline
\endfoot
  \hline
\multicolumn{1}{@{}l}{\rlap{\parbox[t]{1.0\textwidth}{\small
%\multicolumn{4}{@{}l@{}}{\hbox to 0pt{\parbox{85mm}{\footnotesize                                                                                        
\footnotemark[$*$]The heliocentric distance ($R_{\rm h}$), geocentric distance ($\Delta$), and phase angle ($\alpha$) for asteroid observation were obtained from the National Aeronautics and Space Administration (NASA) Jet Propulsion Laboratory (JPL) HORIZONS ephemeris generator system.\footnotemark[2]\\
}}}
\endlastfoot
  \hline
  9950 & ESA                      & 2013.12.08 & 1.558 & 0.577 & 5.6  & KOOLS/188cm  & HD 25680\\
 25143 & Itokawa                  & 2000.12.31 & 1.364 & 0.549 & 37.0 & FOCAS/Subaru & Feige 34\\
163899 & 2003 $\mathrm{SD_{220}}$ & 2006.12.23 & 0.912 & 0.338 & 92.0 & FOCAS/Subaru & HD 107368, HD 103930 \\
414990 & 2011 $\mathrm{EM_{51}}$  & 2014.02.22 & 1.387 & 0.414 & 13.4 & FOCAS/Subaru & SA 102-1081\\
       & 2013 NJ                  & 2013.12.06 & 1.022 & 0.042 & 29.9 & KOOLS/188cm  & SA 98-978\\
\end{longtable}
\footnotetext[2]{$\langle$http://ssd.jpl.nasa.gov/horizons.cgi\#top$\rangle$.}
%%%%%%%%%%%%%%%%%%%%%%%%%%%%%%%%%%%%%%%

Before August 2001, between September 2001 and May 2010, and after June 2010, the FOCAS instrument detector had two 2048 $\times$ 4096 SITe ST-002A charge-coupled devices (CCDs), two 2048 $\times$ 4096 Massachusetts Institute of Technology/Lincoln Laboratory CCDs, and two 2048 $\times$ 4176 Hamamatsu Photonics CCDs, respectively.
The three generations of FOCAS CCDs are of the same size, with 15 $\mu$m square pixels, and provide a \timeform{6'} circular field-of-view, with a pixel scale of \timeform{0''.1}.
The 2048 $\times$ 4096 SITe ST-002A CCD for KOOLS has 15 $\mu$m square pixels, giving a \timeform{5'} $\times$ \timeform{4'.4} field-of-view with a pixel scale of \timeform{0''.3}.

Grisms with 6500-, 7500-, and 5500-\AA/mm blaze and SY47 and L600, SO58, and SY47 order-sorting filters for FOCAS spectroscopy were used for the 25143 Itokawa, 163899 2003 $\mathrm{SD_{220}}$, and 414990 2011 $\mathrm{EM_{51}}$ asteroid observations, respectively.
For KOOLS spectroscopy, grisms with 6563-\AA/mm blaze and a Y49 order-sorting filter were utilized.

The FOCAS and KOOLS slit lengths are \timeform{6'.0} and \timeform{4'.4} in the cross-wavelength direction, respectively.
Note that attention must be paid to the slit width selection as atmospheric differential refraction causes erroneous classification of asteroid spectral types.
The atmospheric dispersion corrector (ADC) installed in the Cassegrain unit of the Subaru Telescope can prevent flux loss errors due to atmospheric dispersion.
Observations of Itokawa, 2003 $\mathrm{SD_{220}}$, and 2011 $\mathrm{EM_{51}}$ were performed using slit widths of \timeform{0''.8}, \timeform{1''.0}, and \timeform{4''.0}, respectively.
Considering the orbital element ambiguity and guide error, a slightly wider slit width was employed for observation of 2011 $\mathrm{EM_{51}}$.
As the 1.88-m telescope did not have ADC, it was necessary to select a wide slit for KOOLS observation; thus, a large width of \timeform{6''.0} was utilized for KOOLS.

Non-sidereal tracking was employed for the telescopes tracking the asteroids, except for the Itokawa observations, for which the data was obtained in sidereal tracking mode. 
In that case, the slit length direction was set as the Itokawa movement direction.

Wavelength calibration frames for FOCAS and KOOLS were acquired regularly during the night with light from thorium-argon and iron-neon-argon hollow cathode lamps, respectively.
The reflected spectrum of the target asteroid was obtained by dividing the asteroid spectrum by the solar spectrum.
Hence, division by the spectrum of the observed solar analogue star having the same airmass as the asteroid yielded the reflected spectrum.
This is the standard data acquisition method of asteroidal spectroscopy (e.g., \cite{Bus2002a}).
Observations between the asteroids (except Itokawa) and solar analogue stars were coordinated such that the airmass difference was less than 0.1 in each case.
For Itokawa only, standard stars were observed using the spectroscopic procedures recommended for stars and galaxies.
Stars with spectral models were used as standard stars for the Itokawa observation, and airmass corrections were made using the atmospheric extinction coefficients.
By dividing the acquired asteroid spectrum by the solar spectrum model, the Itokawa reflection spectrum was obtained.

All data reduction for spectroscopy was performed using image reduction and analysis facilities (IRAF).
A collection of flat-field images was acquired by taking a flat-field image each night during the observation period.
The bias was subtracted from all spectral data. 
After bias subtraction, each object frame was divided by a normalized bias to correct the flat-fielding flame.
The extraction of a one-dimensional spectrum from the two-dimensional images was performed using the $apall$ function of the IRAF software.
The obtained asteroid reflectance was normalized at 0.55 \micron.

\subsection{Photometric observations}
Photometric observations of asteroids for colorimetry and lightcurve analyses were recorded using eight telescopes.
The nightly observation details for the colorimetry and lightcurve data are summarized in tables \ref{tab:Colorimetric circumstances} and \ref{tab:Photometric circumstances}, respectively.

%%%%%%%%%%%%%%%%%%%%%%%%%%%%%%%%%%%%%%%
\begin{longtable}{llllllll}
  \caption{Colorimetric circumstances of the asteroids.}\label{tab:Colorimetric circumstances}
  \hline
    No. & Name & Date      & $R_{\rm h}$\footnotemark[$*$] & $\Delta$\footnotemark[$*$] & $\alpha$\footnotemark[$*$] & Telescope & Filter\\
         &     & [YY.MM.DD]&[au]         & [au]     & [$\timeform{D}$] &    & band\\
\endfirsthead
  \hline
    No. & Name & Date      & $R_{\rm h}$ & $\Delta$ & $\alpha$ & Telescope & Band\\
  \hline
\endhead
  \hline
\endfoot
  \hline
%\multicolumn{1}{@{}l}{\rlap{\parbox[t]{1.0\textwidth}{\small
\multicolumn{4}{@{}l@{}}{\hbox to 0pt{\parbox{85mm}{\footnotesize
\footnotemark[$*$]The heliocentric distance ($R_{\rm h}$), geocentric distance ($\Delta$), and phase angle ($\alpha$) for asteroid observation were obtained from the NASA JPL HORIZON ephemeris generator system.\footnotemark[2]\\
}}}
\endlastfoot
  \hline
  2212 & Hephaistos               & 2007.09.06 & 2.342 & 1.745 & 23.2 & UH88 & $BVR_{\rm C}I_{\rm C}$\\
  3361 & Orpheus                  & 2005.11.28 & 1.013 & 0.224 & 76.8 & LOT  & $BVR_{\rm C}I_{\rm C}$\\
  5797 & Bivoj                    & 2005.12.24 & 1.089 & 0.243 & 58.4 & Kiso & $BVR_{\rm C}I_{\rm C}$\\
       &                          & 2005.12.28 & 1.078 & 0.236 & 60.4 & & \\
 11284 & Belenus                  & 2005.11.25 & 1.204 & 0.292 & 37.1 & LOT  & $BVR_{\rm C}I_{\rm C}$\\
       &                          & 2005.11.27 & 1.195 & 0.289 & 38.8 & & \\
 65679 & 1989 UQ                  & 2003.09.26 & 1.156 & 0.202 & 37.2 & Kiso & $BVR_{\rm C}I_{\rm C}$\\
       &                          & 2003.09.29 & 1.157 & 0.193 & 33.6 & & \\
       &                          & 2003.09.30 & 1.157 & 0.189 & 32.3 & & \\
 68278 & 2001 $\mathrm{FC_{7}}$   & 2006.02.04 & 1.588 & 0.605 &  3.6 & Kiso & $BVR_{\rm C}I_{\rm C}$\\
       &                          & 2006.02.05 & 1.589 & 0.605 &  4.1 & & \\
 85585 & Mjolnir                  & 2003.09.29 & 1.128 & 0.134 & 18.9 & Kiso & $BVR_{\rm C}I_{\rm C}$\\
       &                          & 2003.09.30 & 1.123 & 0.129 & 18.7 & &\\
136617 & 1994 CC                  & 2007.12.03 & 1.852 & 0.883 &  8.1 & LOT  & $BVR_{\rm C}$\\
       &                          & 2007.12.06 & 1.866 & 0.907 & 10.0 & & \\
136618 & 1994 $\mathrm{CN_{2}}$   & 2008.02.27 & 2.193 & 1.239 &  9.3 & LOT  & $BVR_{\rm C}I_{\rm C}$\\
       &                          & 2008.02.28 & 2.193 & 1.234 &  8.7 & & \\
       &                          & 2008.04.05 & 2.167 & 1.247 & 13.7 & & \\
137799 & 1999 YB                  & 2005.11.25 & 1.414 & 0.434 &  8.4 & LOT  & $BVR_{\rm C}I_{\rm C}$\\
       &                          & 2005.11.26 & 1.414 & 0.434 &  8.8 & & \\
       &                          & 2005.11.27 & 1.413 & 0.435 &  9.3 & & \\
138404 & 2000 $\mathrm{HA_{24}}$  & 2006.04.24 & 1.128 & 0.319 & 59.7 & Kiso & $BVR_{\rm C}I_{\rm C}$\\
141018 & 2001 $\mathrm{WC_{47}}$  & 2006.12.18 & 1.382 & 0.412 & 12.2 & LOT  & $BVR_{\rm C}I_{\rm C}$\\
       &                          & 2006.12.19 & 1.378 & 0.405 & 11.2 & & \\
141424 & 2002 CD                  & 2006.03.28 & 1.104 & 0.209 & 54.7 & LOT  & $BVR_{\rm C}I_{\rm C}$\\
142348 & 2002 $\mathrm{RX_{211}}$ & 2005.12.24 & 1.155 & 0.200 & 28.1 & Kiso & $BV_{\rm C}RI_{\rm C}$\\
       &                          & 2005.12.28 & 1.167 & 0.213 & 27.7 & & \\
153591 & 2001 $\mathrm{SN_{263}}$ & 2008.02.28 & 1.057 & 0.075 & 26.4 & LOT  & $BVR_{\rm C}I_{\rm C}$\\
154007 & 2002 BY                  & 2007.02.21 & 1.499 & 0.532 & 13.2 & Kiso & $BVR_{\rm C}I_{\rm C}$\\
159467 & 2000 $\mathrm{QK_{25}}$  & 2005.11.25 & 1.418 & 0.477 & 20.8 & LOT  & $BVR_{\rm C}I_{\rm C}$\\
       &                          & 2005.11.28 & 1.427 & 0.478 & 18.6 & & \\
162173 & Ryugu                    & 2007.07.22 & 1.364 & 0.493 & 37.2 & LOT  & $BVR_{\rm C}I_{\rm C}$\\
       &                          & 2007.09.04 & 1.270 & 0.287 & 21.4 & UH88 & $BVR_{\rm C}I_{\rm C}$\\
       &                          & 2007.09.12 & 1.248 & 0.269 & 23.1 & Steward & $BVR_{\rm C}I_{\rm C}$\\
       &                          & 2007.09.13 & 1.246 & 0.268 & 23.6 & & \\
       &                          & 2007.09.14 & 1.243 & 0.266 & 24.1 & & \\
163692 & 2003 $\mathrm{CY_{18}}$  & 2007.12.06 & 1.814 & 0.837 &  6.0 & LOT  & $BVR_{\rm C}$\\
       &                          & 2007.12.07 & 1.818 & 0.842 &  6.2 & & \\
164202 & 2004 EW                  & 2006.03.28 & 1.188 & 0.198 & 14.7 & LOT  & $BVR_{\rm C}I_{\rm C}$\\
       &                          & 2006.03.29 & 1.191 & 0.202 & 15.9 & & \\
163899 & 2003 $\mathrm{SD_{220}}$ & 2006.12.21 & 0.918 & 0.337 & 91.2 & LOT  & $BVR_{\rm C}I_{\rm C}$\\
       &                          & 2006.12.22 & 0.914 & 0.337 & 91.7 & & \\
170891 & 2004 $\mathrm{TY_{16}}$  & 2008.02.08 & 1.193 & 0.241 & 28.1 & Kiso & $BVR_{\rm C}I_{\rm C}$\\
171819 & 2001 $\mathrm{FZ_{6}}$   & 2007.12.07 & 1.620 & 0.684 & 16.7 & LOT  & $BVR_{\rm C}I_{\rm C}$\\
       &                          & 2007.12.08 & 1.621 & 0.683 & 16.3 & & \\
172974 & 2005 $\mathrm{YW_{55}}$  & 2007.12.08 & 1.402 & 0.441 & 15.8 & LOT  & $BVR_{\rm C}$\\
206378 & 2003 RB                  & 2003.09.28 & 1.003 & 0.093 & 86.8 & Kiso & $BVR_{\rm C}I_{\rm C}$\\
       &                          & 2003.09.29 & 1.003 & 0.093 & 86.4 & & \\
       &                          & 2003.09.30 & 1.004 & 0.094 & 86.2 & & \\
363505 & 2003 $\mathrm{UC_{20}}$  & 2003.12.03 & 1.044 & 0.083 & 44.0 & Kiso  & $BVR_{\rm C}I_{\rm C}$\\
       &                          & 2003.12.04 & 1.043 & 0.083 & 44.5 & & \\
       &                          & 2005.11.26 & 1.005 & 0.323 & 77.4 & LOT  & \\
481394 & 2006 $\mathrm{SF_{6}}$   & 2007.11.08 & 1.019 & 0.135 & 74.2 & Kiso & $BVR_{\rm C}I_{\rm C}$\\
       & 2001 $\mathrm{QC_{34}}$  & 2007.09.06 & 1.279 & 0.296 & 20.7 & UH88 & $BVR_{\rm C}I_{\rm C}$\\
       & 2004 $\mathrm{DK_{1}}$   & 2004.04.11 & 1.088 & 0.086 &  3.4 & Kiso & $VR_{\rm C}I_{\rm C}$\\
       & 2004 $\mathrm{XL_{14}}$  & 2006.12.19 & 1.010 & 0.028 & 22.5 & LOT  & $BVR_{\rm C}I_{\rm C}$\\
       & 2005 TF                  & 2005.11.28 & 1.071 & 0.164 & 55.2 & LOT  & $BVR_{\rm C}I_{\rm C}$\\
       & 2006 GB                  & 2006.04.25 & 1.074 & 0.195 & 64.6 & Kiso & $BVR_{\rm C}$\\
       & 2007 $\mathrm{BB_{50}}$  & 2007.02.14 & 1.337 & 0.351 &  4.5 & UH88 & $BVR_{\rm C}I_{\rm C}$\\
       & 2007 $\mathrm{BJ_{29}}$  & 2007.02.14 & 1.390 & 0.564 & 35.6 & UH88 & $BVR_{\rm C}I_{\rm C}$\\
       & 2007 $\mathrm{RV_{9}}$   & 2008.02.28 & 1.081 & 0.132 & 44.1 & LOT  & $BVR_{\rm C}I_{\rm C}$\\
       & 2007 $\mathrm{TU_{24}}$  & 2008.02.08 & 1.030 & 0.056 & 37.5 & Kiso & $BR_{\rm C}I_{\rm C}$\\
       & 2010 $\mathrm{JV_{34}}$  & 2010.05.16 & 1.089 & 0.082 & 18.7 & Kiso & $BVR_{\rm C}I_{\rm C}$\\
       &                          & 2010.05.17 & 1.080 & 0.075 & 22.5 & & \\
       & 2010 $\mathrm{TC_{55}}$  & 2010.11.11 & 1.101 & 0.122 & 22.9 & Kiso & $BVR_{\rm C}I_{\rm C}$\\
  \hline
   852 & Wlandinena               & 2005.12.28 & 3.009 & 2.327 & 15.4 & Kiso & $BVR_{\rm C}I_{\rm C}$\\
  7096 & Napier                   & 2007.02.14 & 1.936 & 1.055 & 17.9 & UH88 & $BVR_{\rm C}I_{\rm C}$\\
 22104 & 2000 $\mathrm{LN_{19}}$  & 2013.10.23 & 1.961 & 1.014 & 12.3 & Pirika & $g'r'i'$\\
       & 2006 $\mathrm{EX_{52}}$  & 2006.12.19 & 2.615 & 1.700 &  9.9 & LOT  & $BVR_{\rm C}I_{\rm C}$\\
       &                          & 2006.12.21 & 2.619 & 1.728 & 11.3 & & \\
       & 2006 $\mathrm{XQ_{56}}$  & 2007.02.14 & 1.448 & 0.575 & 29.4 & UH88 & $BVR_{\rm C}I_{\rm C}$\\

\end{longtable}

%%%%%%%%%%%%%%%%%%%%%%%%%%%%%%%%%%%%%%%                                                                                       
\begin{longtable}{llllllll}
  \caption{Lightcurve observational circumstances of the asteroids.}\label{tab:Photometric circumstances}
  \hline
    No. & Name & Date      & Duration & $R_{\rm h}$\footnotemark[$*$] & $\Delta$\footnotemark[$*$] & $\alpha$\footnotemark[$*$] & Telescope\\
         &     & [YY.MM.DD]& [hr]     &[au]         & [au]     & [$\timeform{D}$] & \\
\endfirsthead
  \hline
    No. & Name & Date      & Duration & $R_{\rm h}$ & $\Delta$ & $\alpha$ & Telescope\\
  \hline
\endhead
  \hline
\endfoot
  \hline
%\multicolumn{1}{@{}l}{\rlap{\parbox[t]{1.0\textwidth}{\small
\multicolumn{4}{@{}l@{}}{\hbox to 0pt{\parbox{85mm}{\footnotesize
\footnotemark[$*$]The heliocentric distance ($R_{\rm h}$), geocentric distance ($\Delta$), and phase angle ($\alpha$) for asteroid observation were obtained from the NASA JPL HORIZON ephemeris generator system.\footnotemark[2]\\
}}}
\endlastfoot
  \hline
   433 & Eros                     & 2005.01.08 & 3.54 & 1.134 & 0.420 & 58.6 & MITSuME\\
       &                          & 2005.01.09 & 3.27 & 1.135 & 0.418 & 58.5 & \\
  1943 & Anteros                  & 2007.02.20 & 1.89 & 1.492 & 0.996 & 41.1 & Kiso\\
       &                          & 2007.02.21 & 1.46 & 1.488 & 1.002 & 41.3 & \\
       &                          & 2009.08.17 & 6.56 & 1.450 & 0.503 & 24.1 & \\
       &                          & 2009.08.19 & 0.47 & 1.458 & 0.504 & 22.8 & \\
       &                          & 2009.08.20 & 3.07 & 1.462 & 0.505 & 22.1 & \\
  5797 & Bivoj                    & 2005.12.22 & 1.59 & 1.095 & 0.247 & 57.3 & Kiso\\
       &                          & 2005.12.24 & 2.33 & 1.089 & 0.243 & 58.4 & \\
       &                          & 2005.12.26 & 2.48 & 1.083 & 0.240 & 59.5 & \\
       &                          & 2005.12.28 & 1.54 & 1.078 & 0.236 & 60.4 & \\
       &                          & 2006.02.04 & 0.63 & 1.070 & 0.190 & 59.0 & \\
 11284 & Belenus                  & 2005.11.25 & 2.38 & 1.204 & 0.292 & 37.1 & LOT\\
 14402 & 1991 DB                  & 2009.01.18 & 1.07 & 1.335 & 0.444 & 31.6 & LOT\\
       &                          & 2009.01.20 & 4.75 & 1.323 & 0.424 & 31.1 & \\
       &                          & 2009.01.21 & 6.68 & 1.318 & 0.416 & 30.9 & \\
       &                          & 2009.01.22 & 5.52 & 1.312 & 0.407 & 30.7 & \\
       &                          & 2009.02.25 & 3.97 & 1.139 & 0.162 & 21.4 & Kiso\\
       &                          & 2009.02.28 & 0.83 & 1.125 & 0.147 & 22.3 & \\
       &                          & 2009.03.01 & 6.20 & 1.122 & 0.144 & 22.8 & \\
       &                          & 2009.03.02 & 1.37 & 1.118 & 0.140 & 23.4 & \\
 68278 & 2001 $\mathrm{FC_{7}}$   & 2006.02.04 & 3.96 & 1.588 & 0.605 &  3.6 & Kiso\\
       &                          & 2006.02.05 & 8.10 & 1.589 & 0.605 &  4.1 & \\
 85585 & Mjolnir                  & 2003.09.27 & 4.27 & 1.139 & 0.146 & 19.5 & Kiso\\
       &                          & 2003.09.28 & 5.11 & 1.134 & 0.141 & 19.2 & \\
       &                          & 2003.09.29 & 4.33 & 1.128 & 0.134 & 18.9 & \\
       &                          & 2003.09.30 & 1.58 & 1.123 & 0.129 & 18.7 & \\
 85867 & 1999 $\mathrm{BY_{9}}$   & 2009.01.18 & 5.49 & 1.374 & 0.521 & 33.6 & LOT\\
       &                          & 2009.01.19 & 0.19 & 1.372 & 0.513 & 33.3 & \\
       &                          & 2009.01.20 & 0.56 & 1.369 & 0.507 & 33.1 & \\
       &                          & 2009.01.21 & 3.55 & 1.366 & 0.500 & 32.9 & \\
       &                          & 2009.01.22 & 2.25 & 1.364 & 0.495 & 32.7 & \\
 89136 & 2001 $\mathrm{US_{16}}$  & 2012.10.19 & 5.86 & 1.574 & 0.583 &  6.2 & Nayuta\\
       &                          & 2012.10.20 & 6.03 & 1.576 & 0.587 &  6.9 & \\
136618 & 1994 $\mathrm{CN_{2}}$   & 2006.03.29 & 7.80 & 2.183 & 1.238 & 11.2 & LOT\\
       &                          & 2008.02.26 & 6.22 & 2.194 & 1.244 &  9.8 & \\
137799 & 1999 YB                  & 2008.10.30 & 3.45 & 1.420 & 0.432 &  7.3 & Kiso\\
       &                          & 2008.10.31 & 1.46 & 1.420 & 0.431 &  6.4 & \\
       &                          & 2008.11.01 & 6.84 & 1.420 & 0.431 &  5.8 & \\
       &                          & 2008.11.03 & 4.67 & 1.420 & 0.430 &  4.2 & \\
138175 & 2000 $\mathrm{EE_{104}}$ & 2007.02.20 & 1.46 & 1.258 & 0.284 & 16.4 & Kiso\\
       &                          & 2007.02.21 & 7.31 & 1.256 & 0.283 & 16.9 & \\
138404 & 2000 $\mathrm{HA_{24}}$  & 2006.04.24 & 3.38 & 1.128 & 0.319 & 59.7 & Kiso\\
       &                          & 2006.04.28 & 1.08 & 1.149 & 0.343 & 57.5 & \\
       &                          & 2006.04.29 & 3.04 & 1.153 & 0.349 & 57.0 & \\
141018 & 2001 $\mathrm{WC_{47}}$  & 2006.11.28 & 2.20 & 1.451 & 0.549 & 25.9 & Kiso\\
142348 & 2002 $\mathrm{RX_{211}}$ & 2005.12.22 & 4.00 & 1.150 & 0.194 & 28.3 & Kiso\\
       &                          & 2005.12.24 & 1.59 & 1.155 & 0.200 & 28.1 & \\
       &                          & 2005.12.26 & 3.41 & 1.162 & 0.207 & 27.8 & \\
       &                          & 2005.12.28 & 1.92 & 1.167 & 0.213 & 27.7 & \\
152560 & 1991 BN                  & 2009.12.09 & 4.14 & 1.301 & 0.391 & 30.7 & LOT\\
       &                          & 2009.12.10 & 4.17 & 1.307 & 0.392 & 29.4 & \\
153591 & 2001 $\mathrm{SN_{263}}$ & 2008.02.28 & 0.97 & 1.057 & 0.075 & 26.4 & LOT\\
154007 & 2002 BY                  & 2007.02.16 & 8.42 & 1.521 & 0.568 & 16.1 & Kiso\\
       &                          & 2007.02.19 & 8.07 & 1.508 & 0.547 & 14.5 & \\
       &                          & 2007.02.20 & 4.52 & 1.504 & 0.540 & 13.9 & \\
       &                          & 2007.02.21 & 5.00 & 1.499 & 0.532 & 13.2 & \\
       &                          & 2007.04.11 & 2.78 & 1.294 & 0.358 & 30.5 & Murikabushi\\
       &                          & 2007.04.12 & 1.24 & 1.291 & 0.357 & 31.1 & \\
       &                          & 2007.04.19 & 4.13 & 1.268 & 0.356 & 36.4 & \\
154991 & Vinciguerra              & 2007.01.13 & 5.66 & 1.192 & 0.329 & 44.2 & Kiso\\
       &                          & 2007.01.14 & 4.98 & 1.190 & 0.325 & 44.3 & \\
       &                          & 2007.01.15 & 5.39 & 1.188 & 0.321 & 44.3 & \\
159402 & 1999 $\mathrm{AP_{10}}$  & 2009.09.16 & 1.60 & 1.187 & 0.190 & 16.1 & MITSuME\\
       &                          & 2009.09.17 & 4.85 & 1.180 & 0.184 & 16.9 & \\
       &                          & 2009.09.18 & 4.19 & 1.174 & 0.179 & 17.8 & \\
       &                          & 2009.09.19 & 3.25 & 1.168 & 0.174 & 18.6 & \\
       &                          & 2009.09.20 & 4.73 & 1.162 & 0.169 & 19.5 & \\
       &                          & 2009.10.15 & 6.18 & 1.044 & 0.079 & 52.1 & \\
       &                          & 2009.10.17 & 6.32 & 1.038 & 0.077 & 55.9 & \\
       &                          & 2009.10.18 & 6.13 & 1.035 & 0.077 & 57.8 & \\
       &                          & 2009.10.20 & 6.31 & 1.030 & 0.076 & 61.5 & \\
       &                          & 2009.12.12 & 3.41 & 1.120 & 0.230 & 48.9 & Kiso\\
       &                          & 2009.12.13 & 1.84 & 1.126 & 0.233 & 47.9 & \\
       &                          & 2009.12.14 & 4.42 & 1.132 & 0.237 & 46.8 & \\
159467 & 2000 $\mathrm{QK_{25}}$  & 2005.12.24 & 0.92 & 1.513 & 0.545 & 11.2 & Kiso\\
       &                          & 2005.12.28 & 1.15 & 1.528 & 0.568 & 13.1 & \\
162567 & 2000 $\mathrm{RW_{37}}$  & 2008.02.06 & 3.05 & 0.968 & 0.081 &100.5 & Kiso\\
       &                          & 2008.02.07 & 2.20 & 0.970 & 0.079 & 99.3 & \\
       &                          & 2008.02.08 & 0.12 & 0.973 & 0.077 & 98.0 & \\
163000 & 2001 $\mathrm{SW_{169}}$ & 2008.09.26 & 4.64 & 1.220 & 0.228 & 15.6 & Kiso\\
       &                          & 2008.09.27 & 1.84 & 1.221 & 0.231 & 16.6 & \\
       &                          & 2008.10.30 & 5.13 & 1.246 & 0.353 & 38.0 & \\
       &                          & 2008.11.01 & 6.06 & 1.248 & 0.362 & 38.8 & \\
163899 & 2003 $\mathrm{SD_{220}}$ & 2006.11.28 & 1.38 & 0.974 & 0.359 & 81.4 & Kiso\\
164202 & 2004 EW                  & 2005.03.05 & 1.12 & 1.091 & 0.099 &  1.5 & LOT\\
       &                          & 2005.03.07 & 5.77 & 1.099 & 0.107 &  4.0 & \\
       &                          & 2006.03.28 & 5.94 & 1.188 & 0.198 & 14.7 & \\
       &                          & 2006.03.29 & 6.97 & 1.191 & 0.202 & 15.9 & \\
190491 & 2000 $\mathrm{FJ_{10}}$  & 2014.08.12 & 3.06 & 1.327 & 0.328 & 14.8 & UH88\\
       &                          & 2014.08.13 & 5.99 & 1.323 & 0.326 & 15.2 & \\
       &                          & 2014.08.14 & 4.75 & 1.320 & 0.324 & 15.9 & \\
       &                          & 2014.08.15 & 3.46 & 1.317 & 0.322 & 16.5 & \\
225312 & 1996 $\mathrm{XB_{27}}$  & 2014.08.12 & 2.83 & 1.163 & 0.251 & 48.3 & UH88\\
       &                          & 2014.08.13 & 3.92 & 1.164 & 0.255 & 48.4 & \\
       &                          & 2014.08.14 & 1.08 & 1.165 & 0.258 & 48.6 & \\
       &                          & 2014.08.15 & 3.67 & 1.166 & 0.261 & 48.7 & \\
250697 & 2005 $\mathrm{QY_{151}}$ & 2010.11.10 & 2.73 & 1.065 & 0.205 & 64.0 & Kiso\\
       &                          & 2010.11.11 & 0.67 & 1.056 & 0.203 & 65.7 & \\
253062 & 2002 $\mathrm{TC_{70}}$  & 2013.02.19 & 8.26 & 1.570 & 0.582 &  1.5 & Nayuta\\
       &                          & 2013.02.20 & 5.63 & 1.568 & 0.580 &  1.1 & \\
341843 & 2008 $\mathrm{EV_{5}}$   & 2009.01.22 & 1.73 & 1.028 & 0.081 & 55.5 & LOT\\
       &                          & 2009.03.01 & 2.36 & 1.038 & 0.166 & 69.1 & Kiso\\
       &                          & 2009.03.02 & 1.71 & 1.038 & 0.168 & 69.4 & \\
357439 & 2004 $\mathrm{BL_{86}}$  & 2015.01.28 & 6.68 & 1.000 & 0.018 & 33.7 & MITSuME\\
       &                          & 2015.01.31 & 6.98 & 1.015 & 0.046 & 47.9 & \\
416186 & 2002 $\mathrm{TD_{60}}$  & 2006.11.24 & 2.08 & 1.114 & 0.300 & 57.7 & Kiso\\
451157 & 2009 $\mathrm{SQ_{104}}$ & 2013.05.01 & 5.92 & 1.062 & 0.083 & 47.7 & BSGC\\
       &                          & 2013.05.07 & 4.42 & 1.087 & 0.103 & 38.6 & \\
       &                          & 2013.05.09 & 4.41 & 1.095 & 0.110 & 36.9 & Pirika\\
       &                          & 2013.05.10 & 3.01 & 1.099 & 0.113 & 36.1 & \\
       &                          & 2013.05.17 & 4.15 & 1.130 & 0.145 & 33.0 & Murikabushi\\
       &                          & 2013.05.18 & 3.09 & 1.134 & 0.150 & 32.8 & \\
       & 2004 $\mathrm{QJ_{7}}$   & 2011.11.03 & 4.27 & 1.104 & 0.238 & 56.3 & Kiso\\
       & 2005 $\mathrm{JU_{108}}$ & 2005.08.29 & 4.44 & 1.269 & 0.280 & 19.8 & Kiso\\
       &                          & 2005.08.31 & 0.49 & 1.278 & 0.286 & 17.9 & \\
       & 2005 TF                  & 2005.12.27 & 2.60 & 1.130 & 0.215 & 42.9 & Kiso\\
       &                          & 2005.12.28 & 1.16 & 1.135 & 0.218 & 41.9 & \\ 
       & 2006 GB                  & 2006.04.25 & 1.12 & 1.074 & 0.195 & 64.6 & Kiso\\
       &                          & 2006.04.27 & 1.16 & 1.078 & 0.203 & 64.2 & \\ 
       &                          & 2006.04.28 & 0.48 & 1.081 & 0.208 & 64.0 & \\ 
       &                          & 2006.04.29 & 0.64 & 1.082 & 0.212 & 63.9 & \\ 
       & 2007 $\mathrm{FK_{1}}$   & 2007.05.07 & 6.28 & 1.081 & 0.094 & 38.2 & Murikabushi\\
       &                          & 2007.05.08 & 3.00 & 1.079 & 0.092 & 39.3 & \\
       &                          & 2007.05.09 & 7.59 & 1.076 & 0.089 & 40.6 & \\
       &                          & 2007.05.10 & 6.37 & 1.073 & 0.087 & 41.7 & \\
       &                          & 2007.05.11 & 4.13 & 1.071 & 0.085 & 42.9 & \\
       &                          & 2007.05.12 & 6.56 & 1.068 & 0.083 & 44.2 & \\
       &                          & 2007.05.13 & 5.16 & 1.066 & 0.082 & 45.4 & \\
       & 2010 $\mathrm{JV_{34}}$  & 2010.05.16 & 6.14 & 1.089 & 0.082 & 18.7 & Kiso\\
       &                          & 2010.05.17 & 4.83 & 1.080 & 0.075 & 22.5 & \\
  \hline
11739  & Baton Rouge              & 2007.12.13 & 4.51 & 3.181 & 2.204 &  2.5 & Kiso\\
19483  & 1998 $\mathrm{HA_{116}}$ & 2009.02.26 & 3.11 & 2.141 & 1.223 & 13.8 & Kiso\\
       &                          & 2009.02.28 & 3.17 & 2.142 & 1.242 & 14.7 & \\
\end{longtable}

%%%%%%%%%%%%%%%%%%%%%%%%%%%%%%%%%%%%%%%  
Sixteen colors and 27 asteroid lightcurves were obtained using the 1.05-m Schmidt telescope at the Kiso Observatory, with a 2k CCD camera having an SITe TK2048E detector (2048 $\times$ 2048 pixels with 24 \micron\ square pixels); this yielded an image scale of \timeform{1''.5}/pixel located in the f/3.1 Schmidt focus. 
The CCD field of view was \timeform{51'} $\times$ \timeform{51'}. 
The typical seeing size was \timeform{3''-4''} for this observatory.

Twenty colorimetric and eight lightcurve observations were performed using the 1.02-m telescope at the Lulin Observatory (LOT), with a PI1300B CCD camera equipped with a e2v CCD36-40 detector. The format of the latter was 1340 $\times$ 1300 pixels with 20 \micron\ square pixels, giving an image scale of \timeform{0".52}/pixel located in the f/8.0 Cassegrain focus. 
The CCD field of view was \timeform{12'} $\times$ \timeform{11'}. 
The typical seeing size was \timeform{1''-2''} at this observatory.

Seven colors and two lightcurves were obtained using the University of Hawaii 2.24-m telescope at the Mauna Kea Observatories (UH88).
A Tektronix CCD camera yielded a 2048 $\times$ 2048 pixel CCD with 24 \micron\ square pixels with an image scale of \timeform{0''.22}/pixel and a sky field of \timeform{7'.5} $\times$ \timeform{7'.5} in the f/10 Cassegrain focus. 
The typical seeing size was $\sim$\timeform{0.8''} at this site.

Lightcurves of three NEAs were recorded using the 1.05-m Murikabushi telescope at the Ishigakijima Astrophysical Observatory, equipped with an Apogee U6 camera comprising a Kodak KAF-1001E 1024 $\times$ 1024 pixel detector with 24 \micron\ square pixels in the f/12 Cassegrain focus.
The projected area was \timeform{6'.2} $\times$ \timeform{6'.2}, which corresponded to an angular resolution of \timeform{0''.36}/pixel before 2008.
After 2008, the projected area was changed to  \timeform{12'.3} $\times$ \timeform{12'.3}, corresponding to an angular resolution of \timeform{0''.72}/pixel, for installation of a 0.5-time focal reducer.
The typical seeing size was \timeform{1''-2''} at this observatory.

Three lightcurves were obtained with the 0.50-m telescope at the Okayama Astrophysical Observatory (MITSuME) \citep{Yanagisawa2010}.
The data were recorded with an Apogee U6 camera with a mounted Kodak KAF-1001E 1024 $\times$ 1024 pixel detector with 24 \micron\ square pixels in the f/6.5 Cassegrain focus.
The field of view was \timeform{25.6'} $\times$ \timeform{25.6'}, which corresponds to an angular resoution of \timeform{1''.5}/pixel.
The typical seeing size was \timeform{1''-2''} at this site.

Lightcurves for two NEAs were obtained using the 2.00-m Nayuta telescope at the Nishiharima Astronomical Observatory.
The MINT imager, which was equipped with a 2068 $\times$ 2064-pixel e2V CCD230-42 detector (15 \micron\ square pixels), was attached to the f/12 Cassegrain focus of the telescope.
This system produced image dimensions of \timeform{0''.32}/pixel, yielding a field of view of \timeform{10'.9} $\times$ \timeform{10'.9}.
The typical seeing size was \timeform{1''-2''} at this observatory.

Color and lightcurve observations were performed using the 1.60-m Pirika Telescope at the Nayoro Observatory.
Data were obtained via the NaCS instrument using the Hamamatsu CCD having 2048 $\times$ 1104 pixels with 15 \micron\ square pixels, with an image scale of \timeform{0''.25}/pixel and a sky field of \timeform{8'.4} $\times$ \timeform{4'.5} in the f/12 Nasmyth focus \citep{Nakao2014}.  
The typical seeing size was \timeform{1''-2''} at this observatory.

A colorimetric observation was performed with the 1.54-m Kuiper Telescope at the Steward Observatory.
The Montreal 4K (Mont4K) imager with Fairchild CCD486 detector yielded a format of 4096 $\times$ 4097 (15 \micron\ square pixels) with an image scale of \timeform{0''.14}/pixel and a sky field of \timeform{9'.7} $\times$ \timeform{9'.7} in the f/13.5 Cassegrain focus.
The typical seeing size was \timeform{1''-2''} at this site.

The asteroid lightcurves were measured at the 1.00-m telescope at the Bisei Spaceguard Center (BSGC), which is equipped with a detector consisting of four Hamamatsu Photonic CCD chips with 4096 $\times$ 2048 pixels of 15 \micron\ square pixels, giving an image scale of \timeform{1``.0}/pixel located in the f/3.0 Cassegrain focus.
The CCD field of view was \timeform{12'} $\times$ \timeform{11'} and the field of view of the CCD chip was \timeform{68'} $\times$ \timeform{34'}.
The typical seeing size was \timeform{4''-5''} at this observatory.

The asteroid observations were performed using non-sidereal tracking to increase the asteroid signal level as much as possible.
The standard-star observation was performed with sidereal tracking.
Dark and flat-field images were obtained each observational night.

For most colorimetric observations, a Johnson-Cousins filter system was utilized.
A Sloan Digital Sky Survey (SDSS) filter system was used for part of the observation.
To correct changes in baseline brightness due to asteroid rotation, colorimetric observations were performed by alternately exchanging the reference filter and other filters.
To acquire a high signal level in a short exposure time, the $R_{\rm C}$ band filter was typically used as the reference filter.
Observations were conducted using such a procedure; for example, with the $R_{\rm C}-V-R_{\rm C}-I_{\rm C}-R_{\rm C}-B-R_{\rm C}$ sequence.
Absolute calibration for colorimetry was conducted by observing standard stars each observational night (\cite{Landolt1992}; \cite{Smith2002}). 

Lightcurve observations were performed using an $R_{\rm C}$ or $r'$ band filter.
To acquire the lightcurve from the difference magnitude between the asteroid and comparison stars within the same sky field, most lightcurve observations were not calibrated based on the standard stars.

The dark images for correction were constructed from a median combination of several dark frames. 
Stacked flat-field images for correction were created using a median combination of several flat-field frames. 
All photometric image frames for an individual night were bias-subtracted and flat-fielded using IRAF. 
The fluxes of the asteroids and of the comparison and standard stars were measured through circular apertures with diameters more than three times that of the full-width at half maximum size, using the $APPHOT$ function in the IRAF software.

As most lightcurves were obtained using relative magnitudes, the fluxes of several to dozens of comparison stars in the same frame as each asteroid were used.
After eliminating the variable stars from the comparison stars, the flux difference between the asteroid and the sum of the comparisons was calculated.
Some lightcurve observations were obtained from the $R_{\rm C}$ band sequence for colorimetry analyses.

\section{Data analysis}
\subsection{Periodic analysis}
Before periodic analysis, the one-way light time of each observation was corrected.
Similar to \citet{Hasegawa2014}, the asteroid rotational periods were determined using two methods: the generalised Lomb-Scargle periodogram (GLS) \citep{Zechmeister2009}, which is an extension of the Lomb-Scargle periodogram \citep{Scargle1982}, and a phase dispersion minimization method (PDM) \citep{Stellingwerf1978} with minimization of the sum of the squares of the adjacent data.
First, periodic analysis of the acquired lightcurve was performed with GLS.
Then, periodic analysis was performed with PDM if there were more than two cycles
The rotational periods were finally determined by considering the best result yielded by the two methods.
If there were insufficient data to determine the PDM, the asteroid period was determined with GLS only.
The results of the periodic analysis are presented in Table \ref{tab:Physical properties}.

%%%%%%%%%%%%%%%%%%%%%%%%%%%%%%%%%%%%%%%  
\begin{landscape}
\setlength{\headsep}{60mm} 
\setlength{\textheight}{160mm} 
\setlength{\tabcolsep}{3pt} 
%{
%\tiny
%\scriptsize
%\footnotesize
%\small
\begin{longtable}{lllllllllll}
  \caption{Physical properties of the asteroids.}\label{tab:Physical properties}
  \hline
    No. & Name & Orbital &$\Delta V$\footnotemark[$*$]&$H$\footnotemark[$*$]& Visible & B-D & RP [hr],   & RP [hr],   & Geometric& SC/OC \\
         &     & group   &[km $\mathrm{s^{-1}}$]&[mag]    & tax\footnotemark[$*$]     & tax\footnotemark[$*$] & this study & literature\footnotemark[$*$] & albedo\footnotemark[$*$]   & ratio\footnotemark[$*$]\\
\endfirsthead
  \hline
    No. & Name & Group &$\Delta V$& $H$&Visible & B-D     & RP, this study  & RP, literature & Albedo & SC/OC\\
  \hline
\endhead
  \hline
\endfoot
  \hline
\multicolumn{1}{@{}l}{\rlap{\parbox[t]{1.0\textwidth}{\small
%\multicolumn{4}{@{}l@{}}{\hbox to 0pt{\parbox{85mm}{\footnotesize                                                                                      
\footnotemark[$*$] The delta-{\it v} for transfer from low-Earth orbit to rendezvous was obtained from the NASA Near-Earth Asteroid Delta-V for Spacecraft Rendezvous database.\footnotemark[3]
$H$ is the absolute magnitude taken from the MPC databases.\footnotemark[4]
The character or value in parentheses is that given in the literature.
The bold-type character in braces is the spectral type classified using data in the literature only.
The italic value of the geometric albedo was obtained by combining the absolute magnitude with the size from the radar.
\\
\footnotemark[$\dagger$] As a result of the inferiority of the observed data, the rotational period value is that given in the literature.
}}}
\endlastfoot
  \hline
   433 & Eros                     & Amor   &  6.069 & 11.2 &(S)      & (Sw)       & 5.26 $\pm$ 0.002& ({\bf 5.2703 $\pm$ 0.00002}) & (0.27 $\pm$ 0.01)& (0.28)\\
  1943 & Anteros                  & Amor   &  5.391 & 15.8 &(S)      & (S)       & 2.869\footnotemark[$\dagger$]& ({\bf 2.8692 $\pm$ 0.00006}) & (0.15 $\pm$ 0.08)&--\\
  2212 & Hephaistos               & Apollo & 10.242 & 13.9 &S/D      & {\bf Sq}  & --& (\textgreater20)&(0.19 $\pm$ 0.02)&--\\
  3361 & Orpheus                  & Apollo &  5.541 & 19.0 &V        & (Q)   & --& (3.58) & (0.36 $\pm$ 0.14)&--\\
  5797 & Bivoj                    & Amor   &  5.741 & 18.8 &S/D, \{{\bf Q}\}& --   & 2.706\footnotemark[$\dagger$]& ({\bf 2.706 $\pm$ 0.003}) & --&--\\
  9950 & ESA                      & Amor   &  7.557 & 16.1 &S        & {\bf Sw}  & --& (6.712 $\pm$ 0.005)& (0.10 $\pm$ 0.05)&--\\
 11284 & Belenus                  & Amor   &  5.800 & 18.2 &Q        & {\bf Q}   & $\sim$5.7& ({\bf 5.43 $\pm$ 0.02}) & (0.25 $\pm$ 0.19)&--\\
 14402 & 1991 DB                  & Amor   &  5.948 & 18.7 &(C)      & (Ch)      & {\bf 2.261 $\pm$ 0.00006}& (2.266 $\pm$ 0.0001)  & (0.14 $\pm$ 0.06)&--\\
 25143 & Itokawa                  & Apollo &  4.632 & 19.2 &S        & (Sqw)     & --& (12.1324 $\pm$ 0.0001)& (0.29 $\pm$ 0.03)& (0.27)\\
 65679 & 1989 UQ                  & Aten   &  6.405 & 19.5 &C/B      & {\bf C}   & --& (7.748 $\pm$ 0.001)& (0.03 $\pm$ 0.01)&--\\
 68278 & 2001 $\mathrm{FC_{7}}$   & Amor   &  5.784 & 18.4 &X        & {\bf Xc}  & 4.230\footnotemark[$\dagger$]& ({\bf 4.230 $\pm$ 0.002})& (0.09 $\pm$ 0.02)&--\\
 85585 & Mjolnir                  & Apollo &  5.608 & 21.6 &A/D      & {\bf S}   & {\bf 11.6 $\pm$ 0.04}& -- & --&--\\
 85867 & 1999 $\mathrm{BY_{9}}$   & Amor   &  6.368 & 18.0 &(Q)      &\{{\bf Q}\} & \textgreater 16& ({\bf \textgreater 25})&--&--\\
 89136 & 2001 $\mathrm{US_{16}}$  & Apollo &  4.428 & 20.2 &(Sq/Sr)  & --        & 14.3 $\pm$ 0.3 & ({\bf 14.39}) & --& (0.39)\\
136617 & 1994 CC                  & Apollo &  5.379 & 17.7 &D/S      & {\bf indet.} & --& ({\bf 2.3886 $\pm$ 0.00009})& ({\it $\sim$0.38})& (0.50)\\
136618 & 1994 $\mathrm{CN_{2}}$   & Apollo &  5.159 & 16.7 &{\bf S}/D& --        & {\bf 13.0 $\pm$ 0.0002}& -- & (0.47 $\pm$ 0.23)&--\\
136923 & 1998 $\mathrm{JH_{2}}$   & Amor   &  6.699 & 16.2 &(S)      &\{{\bf S}\} & --& (15.7 $\pm$ 0.1) & --&--\\
137799 & 1999 YB                  & Amor   &  5.889 & 18.5 &S        & {\bf Sq}  & 9.43 $\pm$ 0.02& ({\bf 9.39 $\pm$ 0.05})& (0.31 $\pm$ 0.21)&--\\
138175 & 2000 $\mathrm{EE_{104}}$ & Apollo &  6.537 & 20.3 &--       &\{{\bf Xn}\}& {\bf 7.6 $\pm$ 0.04}& ($\sim$8)& (0.18 $\pm$ 0.05)& (1.1)\\
138404 & 2000 $\mathrm{HA_{24}}$  & Apollo &  5.813 & 19.1 &A/S/D    & {\bf S}   & 3.777\footnotemark[$\dagger$]& ({\bf 3.7768 $\pm$ 0.2})& --&--\\
141018 & 2001 $\mathrm{WC_{47}}$  & Amor   &  4.800 & 18.9 &Q        & {\bf Sq}  & 16.75\footnotemark[$\dagger$]& ({\bf 16.747 $\pm$ 0.006})&--&--\\
141424 & 2002 CD                  & Aten   &  5.602 & 20.5 &B        & {\bf B}   & --& --     & --&--\\ 
142348 & 2002 $\mathrm{RX_{211}}$ & Amor   &  6.332 & 18.1 &S/A      & {\bf Sw}  & 5.067 $\pm$ 0.001& ({\bf 5.0689 $\pm$ 0.006})& --&--\\
152560 & 1991 BN                  & Apollo &  5.560 & 19.1 &(Q)      &\{{\bf Q}\} & {\bf 3.49 $\pm$ 0.04}& --    & (0.38 $\pm$ 0.25)& (0.27)\\
153591 & 2001 $\mathrm{SN_{263}}$ & Amor   &  5.963 & 16.9 &B/C      & {\bf Cb}  & 3.426\footnotemark[$\dagger$]& ({\bf 3.4256 $\pm$ 0.0002})& (0.05 $\pm$ 0.02)& (0.16)\\
154007 & 2002 BY                  & Amor   &  6.069 & 17.9 &X        & {\bf Xc}  & {\bf 14.91 $\pm$ 0.003}& --  & --&--\\
154991 & Vinciguerra              & Amor   &  5.915 & 18.4 &(Q)      & --        & {\bf 3.385 $\pm$ 0.0009}& --  & --&--\\
159402 & 1999 $\mathrm{AP_{10}}$  & Apollo &  6.429 & 16.1 &(Sa)     & (Sq)      & {\bf 7.911 $\pm$ 0.0001}& (7.908 $\pm$ 0.001)& (0.34 $\pm$ 0.17)&--\\
159467 & 2000 $\mathrm{QK_{25}}$  & Amor   &  6.617 & 18.2 &{\bf Q}  & --        & {\bf \textgreater 3}& -- & --&--\\
162173 & Ryugu                    & Apollo &  4.646 & 19.3 &C/B      & (C)       & --& (7.6311)  & (0.05 $\pm$ 0.01)&--\\
162567 & 2000 $\mathrm{RW_{37}}$  & Apollo &  6.154 & 19.9 &(C)      & --        & {\bf 2.439 $\pm$ 0.007}& -- & (0.17 $\pm$ 0.03) &--\\
163000 & 2001 $\mathrm{SW_{169}}$ & Amor   &  5.257 & 19.2 &(S/Sr/Sq)&\{{\bf Sw}\}& 69.97\footnotemark[$\dagger$]& ({\bf 69.97 $\pm$ 0.05})& --&--\\
163692 & 2003 $\mathrm{CY_{18}}$  & Apollo &  5.724 & 18.0 &{\bf X/C}  & --      & --& -- & --& (0.20)\\
163899 & 2003 $\mathrm{SD_{220}}$ & Aten   &  7.855 & 17.3 &S/D, S   & {\bf Sw}  & 285\footnotemark[$\dagger$]& ({\bf $\sim$285})  & (0.31 $\pm$ 0.04)&--\\
164202 & 2004 EW                  & Aten   &  6.730 & 20.8 &C/B      & {\bf Xn}  & {\bf 11.11 $\pm$ 0.0002}& --  & (0.35 $\pm$ 0.18)&--\\
170891 & 2004 $\mathrm{TY_{16}}$  & Amor   &  6.578 & 16.9 &S/D      & {\bf D}   & --& (2.795 $\pm$ 0.002)  & --&--\\
171819 & 2001 $\mathrm{FZ_{6}}$   & Amor   &  6.520 & 18.3 &V        & {\bf V}   & --& (15.24 $\pm$ 0.04)  & --&--\\
172974 & 2005 $\mathrm{YW_{55}}$  & Amor   &  6.379 & 19.3 &{\bf S}/D& --        & --& --     & (0.30 $\pm$ 0.24)&--\\
190491 & 2000 $\mathrm{FJ_{10}}$  & Apollo &  4.553 & 20.9 &(S)      & --        & {\bf 2.456 $\pm$ 0.002}& (\textgreater2) & --&--\\
206378 & 2003 RB                  & Apollo &  5.510 & 18.7 &S/A      & {\bf Sw}  & --& (37.5 $\pm$ 0.2) & (0.44 $\pm$ 0.19)&--\\
214869 & 2007 $\mathrm{PA_{8}}$   & Apollo &  6.797 & 16.4 &(Q/O)    &\{{\bf Sq}\}& --& (102 $\pm$ 5) & (0.29 $\pm$ 0.08)& (0.29)\\
225312 & 1996 $\mathrm{XB_{27}}$  & Amor   &  4.755 & 21.7 &(D), \{{\bf S}/D\}& --   & {\bf 1.195 $\pm$ 0.002}& --  & (0.48 $\pm$ 0.26)&--\\
250697 & 2005 $\mathrm{QY_{151}}$ & Apollo &  7.006 & 17.8 &--       & --        & {\bf \textgreater 11}& --&--&--\\
253062 & 2002 $\mathrm{TC_{70}}$  & Amor   &  4.886 & 21.0 &(Sq/Sr)  & --        & {\bf $\sim$17}& --&--&--\\
341843 & 2008 $\mathrm{EV_{5}}$   & Aten   &  5.633 & 20.0 &(C)      &\{{\bf B}\} & 3.725\footnotemark[$\dagger$]& ({\bf 3.725 $\pm$ 0.001})& (0.10 $\pm$ 0.31)& (0.40)\\
357439 & 2004 $\mathrm{BL_{86}}$  & Apollo &  8.458 & 19.4 &(V)      & (V)       & 2.572 $\pm$ 0.001& ({\bf 2.620})& ({\it $\sim$0.25})&--\\
363505 & 2003 $\mathrm{UC_{20}}$  & Aten   &  8.715 & 18.1 &B/C      & {\bf Cb}  & --& --& (0.03 $\pm$ 0.01)& (0.21)\\
414990 & 2011 $\mathrm{EM_{51}}$  & Apollo &  5.213 & 21.9 &{\bf A/S}& --        & --& -- & --&--\\
416186 & 2002 $\mathrm{TD_{60}}$  & Amor   &  5.373 & 19.3 &(S)      &\{{\bf Sw}\}& 2.851\footnotemark[$\dagger$]& {\bf (2.8513 $\pm$ 0.0001)}& (0.50 $\pm$ 0.35)& (0.41)\\
451157 & 2009 $\mathrm{SQ_{104}}$ & Apollo &  4.879 & 21.0 &(Sq)     &\{{\bf Sq}\}  & {\bf 6.856 $\pm$ 0.0006}& --& --&--\\
471240 & 2011 $\mathrm{BT_{15}}$  & Apollo &  4.975 & 21.7 &(A)      &\{{\bf Sw}\}& --& (0.109138 $\pm$ 0.000002)& --&--\\
481394 & 2006 $\mathrm{SF_{6}}$   & Aten   &  6.802 & 19.9 &{\bf A/S}/D& --   & --& --     & (0.21 $\pm$ 0.15)&--\\
       & 2001 $\mathrm{QC_{34}}$  & Apollo &  4.972 & 20.1 &S/X, \{{\bf Q}\}& --   & --& --     & (0.27 $\pm$ 0.19)&--\\
       & 2004 $\mathrm{DK_{1}}$   & Amor   &  5.627 & 21.0 &{\bf S/D/A}  & --        & --& -- & --&--\\
       & 2004 $\mathrm{QJ_{7}}$   & Apollo &  6.159 & 18.5 &--       & (Q)       & {\bf 1.28 $\pm$ 0.02}& --  & --&--\\
       & 2004 $\mathrm{XL_{14}}$  & Aten   & 12.436 & 21.2 &X/D/S    & {\bf Xk}  & --& --     & (0.15 $\pm$ 0.10)&(0.49)\\
       & 2005 $\mathrm{JU_{108}}$ & Amor   &  8.436 & 19.7 &(C)      & --        & {\bf 5.34 $\pm$ 0.03}& -- & --&--\\
       & 2005 TF                  & Amor   &  5.899 & 20.2 &Q        & {\bf Sqw} & 2.57\footnotemark[$\dagger$]& ({\bf2.57 $\pm$ 0.005}) & --&--\\
       & 2006 GB                  & Aten   &  6.341 & 20.3 &{\bf Q/X/S/D}& --    & {\bf \textgreater 2.5}& -- & --&--\\
       & 2007 $\mathrm{BB_{50}}$  & Amor   &  8.436 & 18.5 &{\bf S/D}& --        & --& -- & --&--\\
       & 2007 $\mathrm{BJ_{29}}$  & Apollo &  9.970 & 18.8 &{\bf Q}  & --        & --& -- & --&--\\
       & 2007 $\mathrm{FK_{1}}$   & Amor   &  6.570 & 20.2 &--       & --        & {\bf 17.11 $\pm$ 0.05}& --  & --&--\\
       & 2007 $\mathrm{RV_{9}}$   & Apollo &  6.624 & 20.2 &{\bf Q/S}& --      & --& -- & --&--\\
       & 2007 $\mathrm{TU_{24}}$  & Apollo &  6.093 & 20.3 &{\bf Q}  & --        & --& ($\sim$26)  & --&(0.37)\\
       & 2010 $\mathrm{JV_{34}}$  & Apollo &  6.754 & 20.8 &{\bf Q}/C& --        & {\bf 2.783}& -- & (0.17 $\pm$ 0.04)&--\\
       & 2010 $\mathrm{TC_{55}}$  & Amor   &  8.350 & 20.3 &{\bf A/S}& --        & --& (2.446 $\pm$ 0.04) & --&--\\
       & 2013 NJ                  & Apollo &  4.908 & 22.0 &V/Q      & {\bf Q}   & --& (2.02 $\pm$ 0.05)  & --&--\\
  \hline
   852 & Wladilena                & Inner MBA &     & 10.0 &{\bf S}  & --        & --& (4.6133 $\pm$ 0.0003)&(0.28 $\pm$ 0.01)&--\\
  7096 & Napier                   & Mars-crosser&   & 15.1 &{\bf C}  & --        & --& --& (0.05 $\pm$ 0.01)&--\\
 11739 & Baton Rouge              & Hilda  &        & 12.0 &{\bf D}       & --        & {\bf 4.8 $\pm$ 0.3}& --  & (0.08 $\pm$ 0.01)&--\\ 
 19483 & 1998 $\mathrm{HA_{116}}$ & Inner MBA  &    & 13.9 &--       & --        & 2.619 $\pm$ 0.004& ({\bf 2.6259 $\pm$ 0.0002}) & --&--\\
 22104 & 2000 $\mathrm{LN_{19}}$  & Middle MBA &    & 13.5 &{\bf X}/C& --        & --& --& (0.13 $\pm$ 0.02)&--\\
       & 2006 $\mathrm{EX_{52}}$  & Damocloid&      & 14.6 &{\bf D}  & --        & --& --& --&--\\
       & 2006 $\mathrm{XQ_{56}}$  & Cybele   &      & 17.9 &{\bf D}  & --        & --& --& (0.03 $\pm$ 0.03)&--\\
\end{longtable}
%}
\end{landscape}
\footnotetext[3]{$\langle$https://echo.jpl.nasa.gov/~lance/delta\_v/delta\_v.rendezvous.html$\rangle$.}
\footnotetext[4]{$\langle$https://minorplanetcenter.net/data$\rangle$.}

%%%%%%%%%%%%%%%%%%%%%%%%%%%%%%%%%%%%%%%  
\subsection{Spectral classification}
The asteroid colorimetric data were converted to reflectance data using the solar color \citep{Ramirez2012}.
Spectral classification of the observed visible spectroscopic and spectrophotometric asteroidal data was performed using the Tholen taxonomy \citep{Tholen1984}.
Although the spectral data contained sufficient wavelength coverage for classification, they were ultimately insufficient for classification purposes. 
This was because the spectrophotometric data discussed in this study incorporated less than five colors.
Therefore, classification based on spectrophotometric data involved selection of possible spectral types within the error range.
The Tholen taxonomy results are listed in Table \ref{tab:Physical properties}.

In addition, asteroid classification in terms of the Bus-DeMeo taxonomy \citep{DeMeo2009} was performed through a combination of the observed visible data and near-infrared data reported in the literature, where classification had not already been performed.
An online tool\footnotemark[5] was used to classify the asteroids according to the updated Bus–DeMeo taxonomy\footnotemark[6]. 
The Bus-DeMeo taxonomy results are listed in Table \ref{tab:Physical properties}.

\footnotetext[5]{$\langle$http://smass.mit.edu/busdemeoclass.html$\rangle$.}
\footnotetext[6]{R. P. Binzel et al. (in preparation) added the Xn-type. See also \citet{Hasegawa2017}.}

\section{Results}
\subsection{Near-Earth asteroids}
In addition to the asteroids observed in this study, information updates from the previous study \citep{Kuroda2014} are also presented below.

\subsubsection{433 Eros}
Asteroid 433 Eros (1898 DQ) belongs to the Amor group, has a delta-$v$ of 6.069 km $\mathrm{s^{-1}}$, and was first rendezvoused with a spacecraft (NEAR-Shoemaker) (e.g., \cite{Yeomans2000}).
Eros is classified as S-type in the Tholen taxonomy \citep{Tholen1984}, the Bus taxonomy \citep{Bus2002}, and the Bus-DeMeo taxonomy \citep{Thomas2014}.
The asteroid albedo is 0.27 \citep{Usui2011} and it has a radar echo circular polarization ratio (SC/OC) of 0.28 \citep{Benner2008}.
\citet{Campa1938} first reported the Eros rotational period, and many studies have since been performed on this property.
The accurate rotational period was reported to be 5.27026 hr with NEAR-Shoemaker \citep{Yeomans2000}.
Lightcurve observations of Eros were conducted during its apparition in January 2005, and a rotational period of 5.26 hr was obtained (Fig. \ref{fig:L1}).
The result obtained in this study is in agreement with the previous report. 

\begin{figure*}
  \begin{center}
    \FigureFile(180mm,180mm){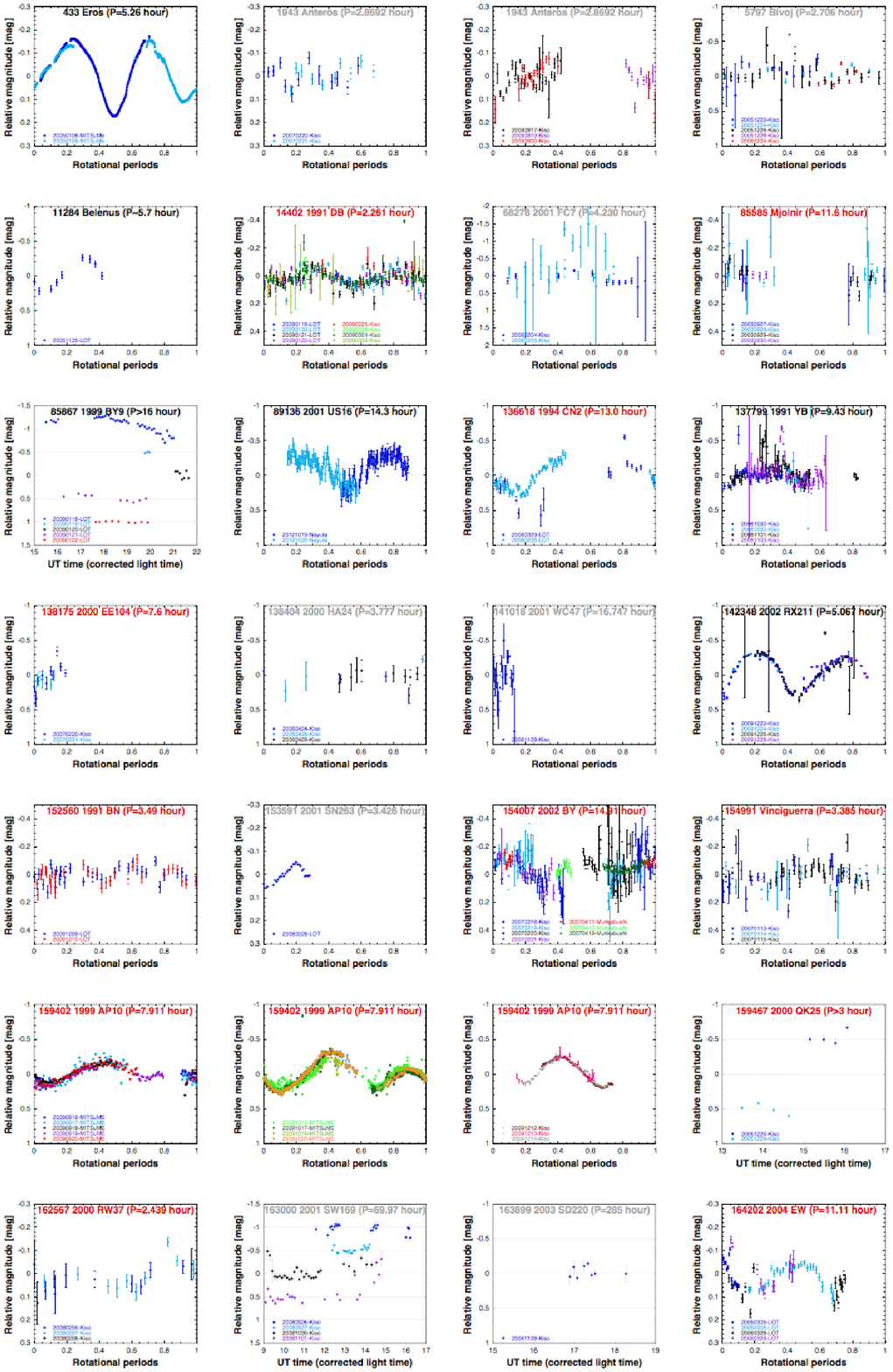}
    %%% \FigureFile(width,height){filename}
  \end{center}
  \caption{Composite rotational lightcurves of asteroids.
}
\label{fig:L1}
\end{figure*}
\addtocounter{figure}{-1}

\begin{figure*}
  \begin{center}
    \FigureFile(180mm,180mm){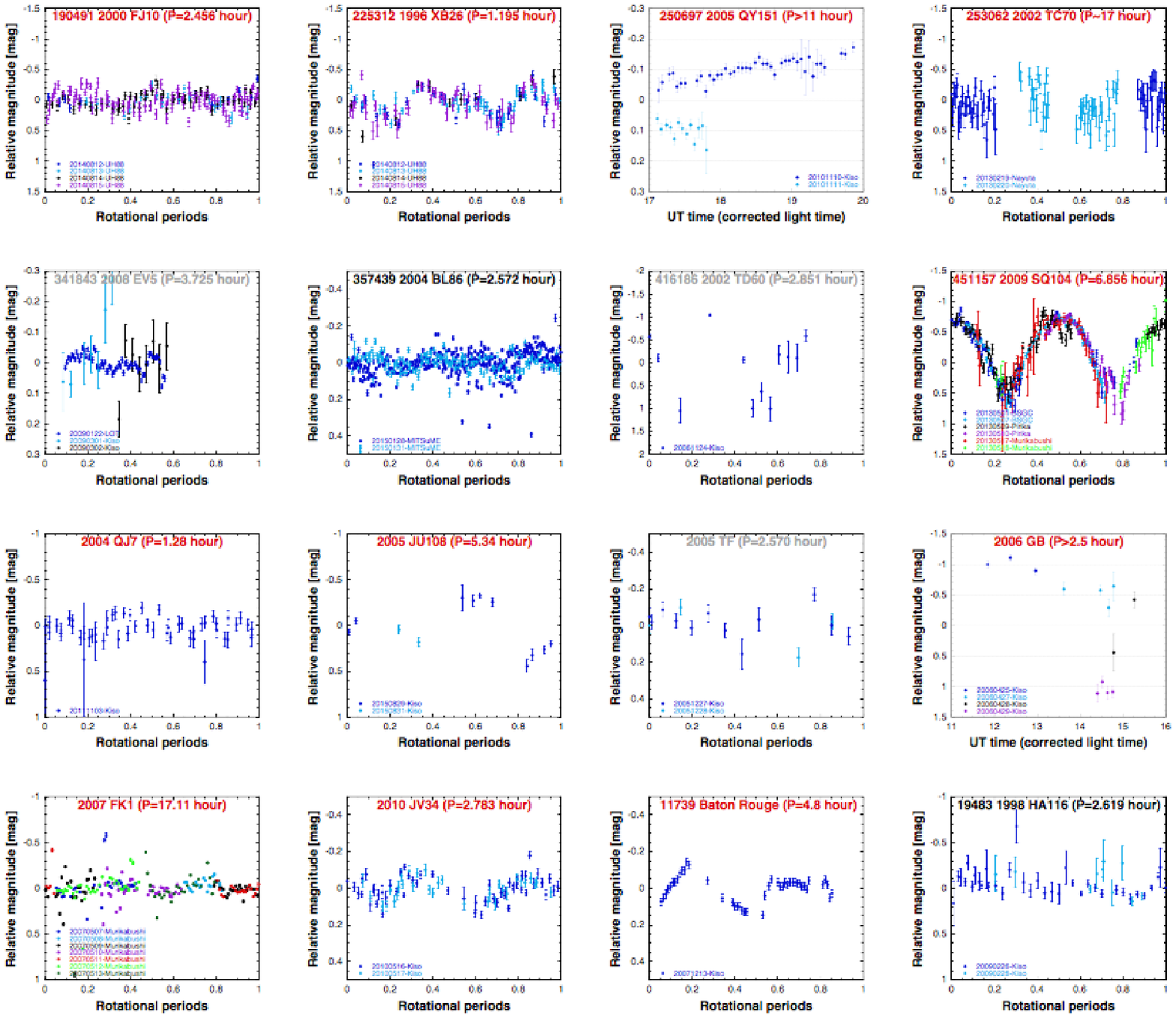}
    %%% \FigureFile(width,height){filename}
  \end{center}
  \caption{(continued)
}
\label{fig:L1}
\end{figure*}

\subsubsection{1943 Anteros}
Asteroid 1943 Anteros (1973 EC) is a member of the Amor group and has a delta-$v$ of 5.391 km $\mathrm{s^{-1}}$.
It was classified as S-type by \citet{Tholen1984}, L-type by \citet{Binzel2001a}, and S-type by \citet{DeMeo2009}.
The albedo of this asteroid has been estimated to be 0.15 \citep{Trilling2010}.
An accurate rotational period of 2.8692 hr has been reported for Anteros \citep{Warner2017c}.
For periodic analysis, Anteros was observed during its apparition in February 2007 and August 2009.
However, the rotational period could not be determined in this study because of the lack of data for periodic analysis.
The value of the rotational period  was quoted from \citet{Warner2017c} (Fig. \ref{fig:L1}).

\subsubsection{2212 Hephaistos}
Asteroid 2212 Hephaistos (1978 SB) is a member of the Apollo group and has a delta-$v$ of 10.242 km $\mathrm{s^{-1}}$.
The albedo is 0.19 \citep{Usui2011} and \citet{Pravec1997} have reported that the asteroid has a rotational period of more than 20 hr.
Visible colorimetric, visible spectroscopic, and near-infrared spectroscopic data of the asteroid are reported in \citet{Pravec1997} and Tedesco 2005\footnotemark[7], \citet{Luu1990}, and \citet{deLeon2010} and \citet{DeMeo2014}.
\citet{DeMeo2014} have demonstrated that Hephaistos is an Sq/Q-type asteroid using near-infrared data only.
Colorimetric observation for Hephaistos was performed during its apparition in September 2007.
The spectrophotometric data in this study allowed classification of this asteroid to S/D-type (Fig. \ref{fig:S1}).
This asteroid was classified as Sq-type in combination with the visible and infrared data.
Several spectra including those from the literature were compared in this work. 
The spectra are roughly coincident, but there are slight differences in shape.

\footnotetext[7]{Tedesco, E. F. 2005, NASA Planetary Data System, EAR-A-5-DDR-UBV-MEAN-VALUES-V1.2.}

\begin{figure*}
  \begin{center}
    \FigureFile(180mm,180mm){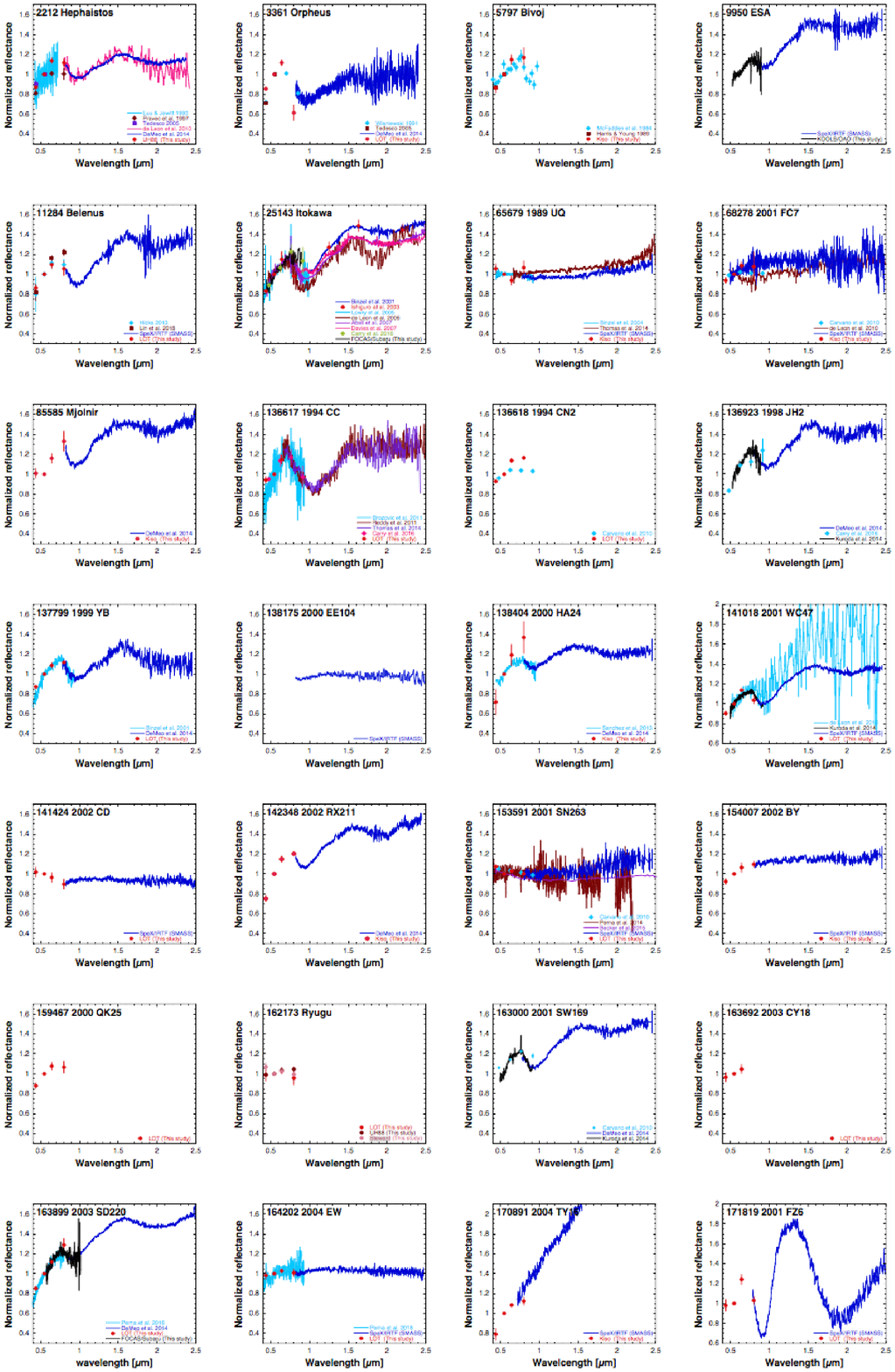}
    %%% \FigureFile(width,height){filename}
  \end{center}
  \caption{spectra of asteroids.
}
\label{fig:S1}
\end{figure*}
\addtocounter{figure}{-1}

\begin{figure*}
  \begin{center}
    \FigureFile(180mm,180mm){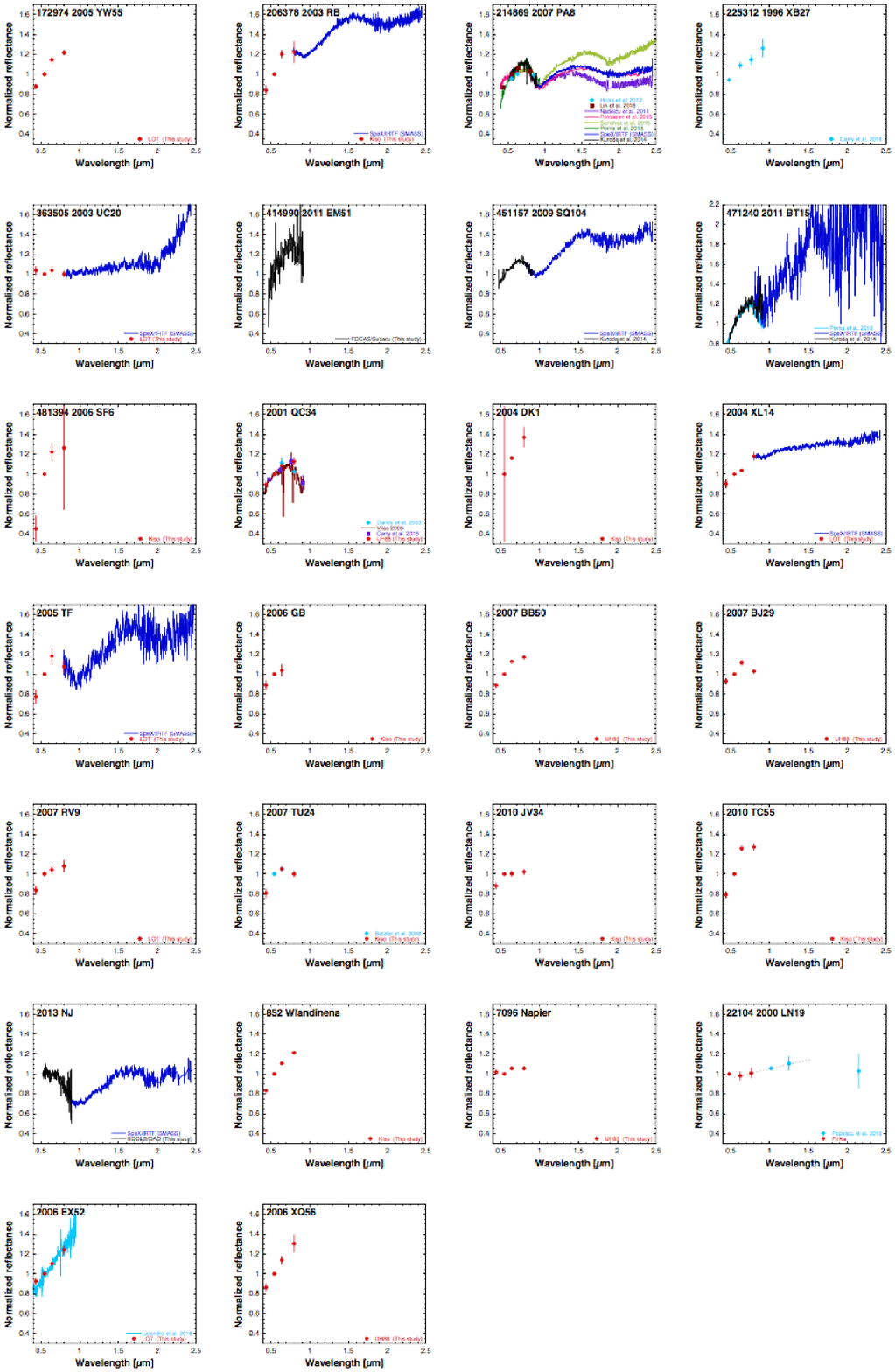}
    %%% \FigureFile(width,height){filename}
  \end{center}
  \caption{(continued)
}
\label{fig:S1}
\end{figure*}

\subsubsection{3361 Orpheus}
Asteroid 3361 Orpheus (1982 HR) has an Apollo-type orbit with a delta-$v$ of 5.541 km $\mathrm{s^{-1}}$.
The geometric albedo is 0.36 \citep{Mainzer2011} and the rotation period is reported as 3.58 hr \citep{Wisniewski1991}.
\citet{Wisniewski1991} and \citet{DeMeo2014} have classified the asteroid as V- and Q-type, respectively.
Colorimetric observation for Orpheus was conducted during its apparition in September 2005.
In this study, the asteroid was classified as V-type using spectrophotometric data (Fig. \ref{fig:S1}).

\subsubsection{5797 Bivoj}
Asteroid 5797 Bivoj (1980 AA) with a delta-$v$ of 5.741 km $\mathrm{s^{-1}}$ belongs to the Amor group.
A rotational period of 2.706 hr has been reported for Bivoj \citep{Mottola1995}.
Photometric observations of this asteroid were performed during its apparition in December 2005 and February 2006.
Because of a lack of lightcurve data to conduct periodic analysis, the rotational period of the asteroid could not be determined in this study.
The value of the rotational period was cited from \citet{Mottola1995} (Fig. \ref{fig:L1}).
Visible colorimetric observations were performed (\cite{Harris1989}; \cite{McFadden1984}).
From the colorimetric data in this study, the asteroid was categorized as S/D-type.
Although classification was not performed by \citet{McFadden1984}, the classification in this work was conducted using the data reported by \citet{McFadden1984}, because the wavelength range considered in that work was wider than that of the present study.
This asteroid was classified as Q-type.

\subsubsection{9950 ESA}
Asteroid 9950 ESA (1990 VB) is a member of the Amor group and has a delta-$v$ of 7.557 km $\mathrm{s^{-1}}$.
\citet{Warner2014a} showed that ESA has a rotational period of 6.712 hr.
Further, the albedo is 0.10 \citep{Trilling2010}.
Spectroscopic observation for ESA was performed during its apparition in December 2013.
The asteroid was classified as S-type using visible data (Fig. \ref{fig:S1}).
In addition, by combining the visible data in this study and the near-infrared data in the Database of The MIT-University of Hawaii-Infrared Telescope Facility (MIT-UH-IRTF) Joint Campaign for Near-Earth Object (NEO) Spectral Reconnaissance (SMASS)\footnotemark[8], the asteroid was classified as Sw-type.

\footnotetext[8]{$\langle$http://smass.mit.edu/minus.html$\rangle$.}

\subsubsection{11284 Belenus}
Asteroid 11284 Belenus (1990 BA), belongs to the Amor group and has a delta-$v$ of 5.800 km $\mathrm{s^{-1}}$.
A rotational period of 5.43 hr was reported by \citet{Hicks2013}.
The NEOSurvey website\footnotemark[9] reports that Belenus has a geometric albedo of 0.25.
Belenus was observed during its apparition in November 2005.
The asteroid rotational period was determined to be 5.7 hr from its lightcurve (Fig. \ref{fig:L1}).
The results of this study are in somewhat good agreement with past reports.
Belenus was classified as S-type (\cite{Hicks2013}; \cite{Lin2018}).
The spectrophotometric data obtained in this study indicate a Q-type asteroid (Fig. \ref{fig:S1}).
A combination of the visible data acquired in this study and near-infrared SMASS data\footnotemark[8] also indicates a Q-type asteroid.

\footnotetext[9]{$\langle$http://nearearthobjects.nau.edu/neosurvey/$\rangle$.}

\subsubsection{14402 1991 DB}
Asteroid 14402 1991 DB with a delta-$v$ of 5.948 km $\mathrm{s^{-1}}$ is assigned to the Amor group.
The asteroid was classified as C-type in the Bus taxonomy \citep{Binzel2004b} and Ch-type in the Bus-DeMeo taxonomy \citep{Thomas2014}. 
The albedo of 1991 DB is 0.14 \citep{Delbo2003} and the rotational period is reported to be 2.266 hr on the Ondrejov Asteroid Photometry Project website\footnotemark[10](Pravec web).
Lightcurve observations of 1991 DB were conducted during its apparition between January and March 2009, and a rotational period of 2.261 hr was obtained (Fig. \ref{fig:L1}). 
The result obtained in this study is in agreement with the previous report.

\footnotetext[10]{$\langle$http://www.asu.cas.cz/~ppravec/neo.htm$\rangle$.}

\subsubsection{25143 Itokawa}
Asteroid 25143 Itokawa (1998 $\mathrm{SF_{36}}$), which has a delta-$v$ of 4.632 km $\mathrm{s^{-1}}$, is a member of the Apollo group.
Itokawa was the first asteroid in which a sample return was conducted by a spacecraft (Hayabusa) \citep{Fujiwara2006}.
The albedo of this asteroid has been estimated to be 0.29 \citep{Mueller2014} and the asteroid has an SC/OC of 0.27 \citep{Benner2008}.
A rotational period of 12.1324 hr has been reported \citep{Nishihara2018}.
Spectroscopic and spectrophotometric data for Itokawa have been reported in \citet{Binzel2001b}, \citet{Ishiguro2003}, \citet{Lowry2005}, \citet{deLeon2006}, \citet{Abell2007}, \citet{Davies2007}, and \citet{Carry2016}, and this asteroid  was classified as Sqw-type in the Bus-DeMeo taxonomy (\cite{Thomas2014}; \cite{Hasegawa2017}).
Itokawa was spectrally observed during its apparition in December 2000. 
The spectroscopic data acquired in this study yielded a classification as S-type (Fig. \ref{fig:S1}).
Several spectra including those from the literature were obtained and compared; the spectra are roughly coincident, but there are slight differences in shape.

\subsubsection{65679 1989 UQ}
Asteroid 65679 1989 UQ has an Aten-type orbit with a delta-$v$ of 6.405 km $\mathrm{s^{-1}}$.
The geometric albedo of the asteroid is 0.03 \citep{Mainzer2012} and the rotation period is reported as 7.748 hr (Pravec web\footnotemark[10]).
The asteroid has been classified as B- and C-type by \citet{Binzel2004a} using visible data and by \citet{Thomas2014} using near-infrared data, respectively. 
Colorimetric observation for 1989 UQ was conducted during its apparition in September 2005.
In this study, the asteroid was classified as a B/C-type asteroid using spectrophotometric data (Fig. \ref{fig:S1}). 
A combination of the visible data and near-infrared SMASS data\footnotemark[8] also indicated a C-type asteroid.
Several spectra including those from the literature were obtained and compared. 
The spectra are roughly coincident, but there are slight differences in its shape.

\subsubsection{68278 2001 $\mathbf{FC_{7}}$}
Asteroid 68278 2001 $\mathrm{FC_{7}}$ is a member of the Amor group and has a delta-v of 5.784 km $\mathrm{s^{-1}}$.
The albedo is 0.09 \citep{Mainzer2012}.
Photometric observations of the asteroid were performed during its apparition in February 2006.
\citet{Monteiro2017} reported that 2001 $\mathrm{FC_{7}}$ has a rotational period of 4.230 hr.
The rotational period, however, could not be determined in this study because there were insufficient data for rotational analysis. 
The spectrophotometric and spectroscopic data of the asteroid are reported in \citet{Carvano2010} and \citet{deLeon2010}.
From the colorimetric data considered in this study, this asteroid was classified as X-type (Fig. \ref{fig:S1}).
Further, this asteroid was classified as Xc-type based on a combination of the visible and infrared data.
Several spectra including those from the literature were obtained and compared. 
The spectra are roughly coincident, but there are slight differences in shape.

\subsubsection{85585 Mjolnir}
Asteroid 85585 Mjolnir (1998 $\mathrm{FG_{s}}$) with a delta-$v$ of 5.608 km $\mathrm{s^{-1}}$ belongs to the Apollo group.
Photometric observations of Mjolnir were performed during its apparition in September 2003.
A rotational period of 11.6 hr was obtained (Fig. \ref{fig:L1}).
From the spectrophotometric data acquired in this study, the asteroid was categorized as A/D-type (Fig. \ref{fig:S1}).
Moreover, based on a combination of the visible data obtained in this study and near-infrared data from SMASS\footnotemark[8], the asteroid was classified as S-type (Fig. \ref{fig:S1}).

\subsubsection{85867 1999 $\mathbf{BY_{9}}$}
Asteroid 85867 1999 $\mathrm{BY_{9}}$ is a member of the Amor group and has a delta-$v$ of 6.368 km $\mathrm{s^{-1}}$.
\citet{Koehn2014} has shown that this asteroid has a rotational period of more than 25 hr.
Lightcurve observations of 1999 $\mathrm{BY_{9}}$ were performed during its apparition in January 2009, and a rotational period of more than 16 hr was obtained (Fig. \ref{fig:L1}). 
As classification of this asteroid was not performed in \citet{deLeon2010}, the classification reported in this study was conducted using their data.
Hence, this asteroid was classified as Q-type (Fig. \ref{fig:S1}).

\subsubsection{89136 2001 $\mathbf{US_{16}}$}
Asteroid 89136 2001 $\mathrm{US_{16}}$ belongs to the Apollo group and has a delta-$v$ of 4.428 km $\mathrm{s^{-1}}$.
The asteroid has an SC/OC of 0.39 \citep{Benner2008} and has been classified as Sq/Sr-type \citet{Kuroda2014}.
The rotational period of 2001 $\mathrm{US_{16}}$ was shown to be 14.39 hr (Pravec web\footnotemark[10]).
The asteroid was observed during its apparition in October 2012, and a rotational period of 14.3 hr was acquired (Fig. \ref{fig:L1}).
The result of this study is consistent with the previous report.

\subsubsection{136617 1994 CC}
Asteroid 136617 1994 CC, with a delta-$v$ of 5.379 km $\mathrm{s^{-1}}$, is a member of the Apollo group.
\citet{Brozovic2011} has reported that this asteroid, which has an SC/OC of 0.50 and a rotational period of 2.3886 hr, is a triple system composed of three bodies of 0.62, 0.11, and 0.08 km in diameter.
Combining the diameter obtained from radar investigation and the absolute magnitude yields a geometric albedo value of 0.38 in this asteroid.
This asteroid was classified as Sq-, Sa-, and S-type by \citet{Brozovic2011}, \citet{Thomas2014}, and \citet{Carry2016}, respectively.
1994 CC was observed during its apparition in February and April 2008.
The asteroid was classified as D/S-type based on spectrophotometric data in this study (Fig. \ref{fig:S1}). 
Based on a combination of the visible and near infrared data, the classification of 1994 CC in terms of the Bus-DeMeo taxonomy was ``indeterminate''.

\subsubsection{136618 1994 $\mathbf{CN_{2}}$}
Asteroid 136618 1994 $\mathrm{CN_{2}}$ is a member of the Apollo group and has a delta-$v$ of 5.159 km $\mathrm{s^{-1}}$.
The albedo of this asteroid is 0.47 \citep{Trilling2016}. 
\citet{Carvano2010} reported that the asteroid is S-type on the surface.
Photometric observations for 1994 $\mathrm{CN_{2}}$ were performed during its apparition in March 2006 and February and April 2008.
A rotational period of 13.0 hr was obtained (Fig. \ref{fig:L1}).
The spectrophotometric data of this study indicated a classification of S/D-type (Fig. \ref{fig:S1}).
S-type asteroids typically have an albedo of more than 0.1 \citep{Hasegawa2017}; therefore, based on its albedo, this asteroid is constrained to S-type.

\subsubsection{136923 1998 $\mathbf{JH_{2}}$}
Asteroid 136923 1998 $\mathrm{JH_{2}}$ has an Amor-type orbit with a delta-$v$ of 6.699 km $\mathrm{s^{-1}}$.
The rotation period is reported to be 15.7 hr \citep{Vaduvescu2017}. 
\citet{Carry2016}, \citet{Kuroda2014}, and \citet{DeMeo2014} classified this asteroid as D-, S-, and Sq/Q-type, respectively.
As visible and near-infrared spectra of the asteroid had already been acquired, the Bus-DeMeo taxonomy was considered based on both datasets in this study.
This asteroid is S-type (Fig. \ref{fig:S1}).

\subsubsection{137799 1999 YB}
Asteroid 137799 1999 YB with a delta-$v$ of 5.889 km $\mathrm{s^{-1}}$ belongs to the Amor group. 
The value of the asteroid albedo is given as 0.31 on the NEOSurvey website\footnotemark[9].
A rotational period of 9.39 hr has been reported for 1999 YB \citep{Warner2015a}.
\citet{Evans2002} showed that 1999 YB is possibly a long-lived asteroid, as it has low eccentricity and inclination and is not planet-crossing between Earth and Mars. 
Photometric observations of the asteroid  were performed during its apparition in September 2005 and October and November 2008.
The asteroid rotational period was determined to be 9.43 hr from its lightcurve (Fig. \ref{fig:L1}). 
The asteroid was previously categorized as Sq-type (\cite{Binzel2001a}; \cite{DeMeo2009}).
From the colorimetric data obtained in this study, the asteroid was determined to be S-type (Fig. \ref{fig:S1}).
The results of this study are in agreement with the existing literature.

\subsubsection{138175 2000 $\mathbf{EE_{104}}$}
Asteroid 138175 2000 $\mathrm{EE_{104}}$ is a member of the Apollo group and has a delta-$v$ of 6.537 km $\mathrm{s^{-1}}$.
A rotational period of $\sim$8 hr, size of $\sim$0.2 km, and SC/OC polarization ratio of 1.1 were obtained using radar observations \citep{Howell2001}.
An albedo of 0.18 for the asteroid was obtained from information on the size and absolute magnitude.
Lightcurve observations for the asteroid were performed during its apparition in February 2007. 
The rotational period of the asteroid was determined to be 7.6 hr from its lightcurve (Fig. \ref{fig:L1}). 
From the near-infrared data in the SMASS database\footnotemark[8], the asteroid was classified as Xn-type (Fig. \ref{fig:S1}).
Previously, \citet{Benner2008} noted that an E-type asteroid has the highest SC/OC ratio.
The values of the circular polarization ratios and geometric albedo for the asteroid are not contradictory to classification of the asteroid as Xn-type.

\subsubsection{138404 2000 $\mathbf{HA_{24}}$}
Asteroid 138404 2000 $\mathrm{HA_{24}}$ is a member of the Apollo group and has a delta-$v$ of 5.813 km $\mathrm{s^{-1}}$.
The rotational period of 3.7768 hr was shown in \citet{Hayes-Gehrke2017}.
Further, 2000 $\mathrm{HA_{24}}$ was observed during its apparition in April 2006.
The rotational period, however, could not be obtained in this study because of a lack of lightcurve data for periodic analysis. 
Instead, the rotational period was quoted from \citet{Hayes-Gehrke2017} (Fig. \ref{fig:L1}).             
The asteroid was classified as S-type with visible \citep{Sanchez2013} and near-infrared spectra \citep{DeMeo2014}.
The spectrophotometric data in this study yielded an asteroid of A/S/D-type (Fig. \ref{fig:S1}).
A combination of the visible and near-infrared data also indicated an S-type asteroid.

\subsubsection{141018 2001 $\mathbf{WC_{47}}$}
Asteroid 141018 2001 $\mathrm{WC_{47}}$ with a delta-$v$ of 4.800 km $\mathrm{s^{-1}}$ is a member of the Amor group.
The rotational period was reported to be 16.747 hr in \citet{Warner2012}.
Photometric observations of 2001 $\mathrm{WC_{47}}$ were performed during its apparition in November and December 2006.
As there were insufficient data for periodic analysis, the rotational period of the asteroid could not be decided in this study.
The value of the rotational period was cited from \citet{Warner2012} (Fig. \ref{fig:L1}).
Spectroscopic observations were performed (\cite{Kuroda2014}; \cite{deLeon2010}).
From the colorimetric data in this study, the asteroid was categorized as Q-type. 
From a combination of the visible data in this study and the near-infrared data in the SMASS database\footnotemark[8], the asteroid was classified as Sq-type.
Several spectra including those from the literature were obtained and compared. 
The spectra are roughly coincident, but there are slight differences in shape.

\subsubsection{141424 2002 CD}
Asteroid 141424 2002 CD, which a delta-$v$ of 5.602 km $\mathrm{s^{-1}}$, is a member of the Apollo group.
This asteroid was colorimetrically observed during its apparition in March 2006.
The spectrophotometric data obtained in this study allowed classification of the asteroid as B-type (Fig. \ref{fig:S1}).
Additionally, the asteroid class was determined to be B-type based on a combination of the visible spectrophotometric data and near-infrared data from SMASS\footnotemark[8].

\subsubsection{142348 2002 $\mathbf{RX_{211}}$}
Asteroid 142348 2002 $\mathrm{RX_{211}}$ has an Amor-type orbit with a delta-$v$ of 6.332 km $\mathrm{s^{-1}}$.
\citet{Higgins2006} reported that 2002 $\mathrm{RX_{211}}$ has a rotational period of 5.0689 hr. 
Photometric observations of 2002 $\mathrm{RX_{211}}$ were conducted during its apparition in December 2005.
A rotational period of 5.067 h was acquired in this study (Fig. \ref{fig:L1}), which is consistent with the literature.
The asteroid was classified as being S-complex \citep{DeMeo2014}.
The colorimetric acquired data in this study allowed classification of this asteroid as S/A-type (Fig. {\ref{fig:S1}}).
A combination of the visible and near-infrared data indicated an Sw-type asteroid.

\subsubsection{152560 1991 BN}
Asteroid 152560 1991 BN is a member of the Amor group and has a delta-$v$ of 5.560 km $\mathrm{s^{-1}}$.
The albedo of the asteroid is 0.38 \citep{Trilling2016} and it has an SC/OC of 0.27 \citep{Benner2008}.
Visible and near-infrared spectroscopic data for this asteroid are reported in \citet{Binzel2004b} and \citet{Lazzarin2005}, respectively.
Through a combination of the visible data and infrared data in the literature, this asteroid was classified as Q-type (Fig. {\ref{fig:S1}}). 
Lightcurve observations for the asteroid were performed during its apparition in December 2009, and a rotational period of 3.49 hr was obtained (Fig. {\ref{fig:L1}}). 

\subsubsection{153591 2001 $\mathbf{SN_{263}}$}
Asteroid 153591 2001 $\mathrm{SN_{263}}$ with a delta-$v$ of 5.963 km $\mathrm{s^{-1}}$ belongs to the Amor group.
The asteroid has been confirmed as a triple system, having three bodies of 2.5, 0.8, and 0.4 km in diameter \citep{Becker2015}.
The SC/OC ratio and geometric albedo of the triple asteroid are 0.16 \citep{Becker2015} and 0.05 \citep{Delbo2011}, respectively.
\citet{Becker2015}showed that the asteroid has a rotational period of 3.4256 hr.
Photometric observations were performed during its apparition in February 2008.
The rotational period, however, could not be determined in this study because of the short time period spanned by the data for periodic analysis. 
The value of the rotational period was taken from \citet{Becker2015}(Fig. {\ref{fig:L1}). 
2001 $\mathrm{SN_{263}}$ has been classified as B-type (\cite{Carvano2010}; \cite{Perna2014}; \cite{Becker2015}).
From the spectrophotometric data obtained in this study, the asteroid was categorized as B/C-type (Fig. {\ref{fig:S1}}).
Moreover, the combination of the visible data acquired in this study and near-infrared data from the SMASS database\footnotemark[8] allowed classification of the asteroid as Cb-type.
Several spectra including those from the literature were obtained and compared.
These spectra are roughly coincident, but there are slight differences in its shape.

\subsubsection{154007 2002 BY}
Asteroid 154007 2002 BY is a member of the Amor group and has a delta-$v$ of 6.069 km $\mathrm{s^{-1}}$.
Photometric observations of the asteroid were performed during its apparition in February and April 2007.
A rotational period of 14.91 hr was obtained (Fig. {\ref{fig:L1}}).
This asteroid was classified as X-type using visible spectrophotometric data (Fig. {\ref{fig:S1}}).
From a combination of the visible data and infrared data from SMASS\footnotemark[8], the asteroid was categorized as Xc-type.

\subsubsection{154991 Vinciguerra}
Asteroid 154991 Vinciguerra (2005 $\mathrm{BX_{26}}$) belongs to the Amor group and has a delta-$v$ of 5.915 km $\mathrm{s^{-1}}$.
This asteroid was in the Q taxonomic class \citep{Carry2016}.
Vinciguerra was observed during its apparition in January 2009, with a rotational period of 3.385 hr being acquired (Fig. {\ref{fig:L1}}).

\subsubsection{159402 1999 $\mathbf{AP_{10}}$}
Asteroid {159402 1999 $\mathrm{AP_{10}}$, belongs to the Apollo group and has a delta-$v$ of 6.429 km $\mathrm{s^{-1}}$.
1999 $\mathrm{AP_{10}}$ was classified as S-type in the Tholen taxonomy \citep{Ye2011}, Sa-type in the Bus taxonomy \citep{Hicks2009}, and Sq-type in the Bus-DeMeo taxonomy \citep{Thomas2014}.
The asteroid albedo was 0.34 \citep{Trilling2010}. 
The accurate rotational period was reported to be 7.908 hr \citep{Franco2010}.
Lightcurve observations of the asteroid were conducted during its apparition in September, October, and December 2009, and a rotational period of 7.911 hr was obtained (Fig. {\ref{fig:L1}}). 
The result obtained in this study is agreement with the previous report.

\subsubsection{159467 2000 $\mathbf{QK_{25}}$}
The asteroid 159467 2000 $\mathrm{QK_{25}}$ is a member of the Amor group and has a delta-$v$ of 6.617 km $\mathrm{s^{-1}}$.
For periodic analysis, the asteroid was observed during its apparition in December 2005, and a rotational period of more than 2.5 hr was obtained (Fig. {\ref{fig:L1}}). 
Colorimetric observations for the asteroid were conducted during its apparition in November 2005.
The asteroid was classified as Q-type using spectrophotometric data in this study (Fig. {\ref{fig:S1}}). 

\subsubsection{162173 Ryugu}
Asteroid 162173 Ryugu (1999 $\mathrm{JU_{3}}$) is a member of the Apollo group and has a delta-$v$ of 4.646 km $\mathrm{s^{-1}}$.
This asteroid is the target body of the Hayabusa2 mission \citep{Tsuda2013}.
The albedo and rotational period of this asteroid have been estimated to be 0.05 and 7.6311 hr, respectively \citep{Mueller2017}.
The spectroscopic data of the asteroid are reported in \citet{Binzel2001a}, \citet{Vilas2008}, \citet{Lazzaro2013}, \citet{Moskovitz2013}, \citet{Pinilla-Alonso2013}, and \citet{Perna2016}, and indicate that Ryugu is a C-complex asteroid.
Colorimetric observations for the asteroid were performed during its apparition in July and September 2007.
The spectrophotometric data obtained in this study allowed classification of this asteroid as C/B-type (Fig. \ref{fig:S1}).

\subsubsection{162567 2000 $\mathbf{RW_{37}}$}
Asteroid 162567 2000 $\mathrm{RW_{37}}$ has an Apollo-type orbit with a delta-$v$ of 6.154 km $\mathrm{s^{-1}}$.
The geometric albedo of the asteroid is 0.17 \citep{Nugent2015}. 
\citet{Binzel2004b} classified the asteroid as C-type.
Lightcurve observations for 2000 $\mathrm{RW_{37}}$ were conducted during its apparition in February 2008.
A rotational period of 2.439 h was acquired (Fig. {\ref{fig:L1}}). 

\subsubsection{163000 2001 $\mathbf{SW_{169}}$}
Asteroid 163000 2001 $\mathrm{SW_{169}}$ with a delta-$v$ of 5.257 km $\mathrm{s^{-1}}$ belongs to the Amor group.
A rotational period of 69.97 hr was reported for 2001 $\mathrm{SW_{169}}$ \citep{Stephens2016}.
An asteroid with low eccentricity and inclination that is not planet-crossing between the Earth and Mars indicates possible long life \citep{Evans2002}. 
Lightcurve observations of the asteroid were performed during its apparition in September, October, and November 2008.
The long duration lightcurve data for the asteroid were insufficient; thus, the rotational period of the asteroid could not be decided in this study (Fig. {\ref{fig:L1}}).
\citet{Carvano2010}, \citet{Kuroda2014}, and \cite{DeMeo2014} classified the asteroid as S-, S/Sr/Sq-type, and S-complex, respectively.
In combination with the visible data and infrared data in the literature, 2001 $\mathrm{SW_{169}}$ was classified as Sw-type (Fig. \ref{fig:S1}).

\subsubsection{163692 2003 $\mathbf{CY_{18}}$}
Asteroid 163692 2003 $\mathrm{CY_{18}}$ is a member of the Apollo group and has a delta-$v$ of 5.724 km $\mathrm{s^{-1}}$.
The asteroid has an SC/OC of 0.20 \citep{Benner2008}.
Colorimetric observations for 2003 $\mathrm{CY_{18}}$ was performed during its apparition in December 2007. 
The asteroid was classified as X/C-type using visible data (Fig. \ref{fig:S1}).

\subsubsection{163899 2003 $\mathbf{SD_{220}}$}
Asteroid 163899 2003 $\mathrm{SD_{220}}$ belongs to the Aten group and has a delta-$v$ of 7.855 km $\mathrm{s^{-1}}$.
It is reported that the asteroid has a geometric albedo of 0.31 in \citet{Nugent2016}.
A rotational period of 285 hr has been  reported \citep{Warner2016}.
2003 $\mathrm{SD_{220}}$ was observed during its apparition in November and December 2006. 
However, the rotational period could not be determined in this study because of the lack of data for periodic analysis (Fig. \ref{fig:L1}).
2003 $\mathrm{SD_{220}}$ was classified as S/Sr- and Sr-type (\cite{DeMeo2014}; \cite{Perna2016}).
The spectrophotometric and spectroscopic data acquired in this study yielded asteroid classifications of S/D- and S-type, respectively (Fig. \ref{fig:S1}).
A combination of the visible data acquired in this study and the near-infrared data obtained in \citet{DeMeo2014} also indicated Sw-type.

\subsubsection{164202 2004 EW}
Asteroid 164202 2004 EW  with a delta-$v$ of 6.730 km $\mathrm{s^{-1}}$ is classified as the Aten group.
The albedo of 2004 EW is 0.35 \citep{Trilling2010}.
The asteroid was classified as being X-type \citep{Perna2018}.
Photometric observations of 2004 EW were conducted during its apparition in March 2005 and 2006.
The spectrophotometric data of this study yielded classification of the asteroid as C/B-type (Fig. \ref{fig:S1}).
A combination of the visible data acquired in this study and the near-infrared data of the SMASS database\footnotemark[8] allowed classification of the asteroid as Xn-type.
A rotational period of 11.11 hr was obtained (Fig. \ref{fig:L1}).

\subsubsection{170891 2004 $\mathbf{TY_{16}}$} 
Asteroid 170891 2004 $\mathrm{TY_{16}}$, which has a delta-$v$ 6.578 km $\mathrm{s^{-1}}$, is a member of the Amor group. 
The rotational period of this asteroid has been estimated to be 2.795 hr \citep{Carbognani2008}.
Using the colorimetric data acquired in this study, the asteroid was categorized as S/D-type (Fig. \ref{fig:S1}). 
A combination of the visible data and near-infrared SMASS data\footnotemark[8] indicated a D-type asteroid.

\subsubsection{171819 2001 $\mathbf{FZ_{6}}$} 
Asteroid 2001 $\mathrm{FZ_{6}}$ has an Amor-type orbit with a delta-$v$ of 6.520 km $\mathrm{s^{-1}}$.
The rotation period has been reported as 15.24 hr \citep{Warner2017b}.
Colorimetric observations for 2001 $\mathrm{FZ_{6}}$ were conducted during its apparition in December 2007.
The asteroid was classified as V-type using the spectrophotometric data obtained in this study (Fig. \ref{fig:S1}).
A combination of the acquired visible data and the near-infrared SMASS data\footnotemark[8] also indicated a V-type asteroid. 

\subsubsection{172974 2005 $\mathbf{YW_{55}}$} 
Asteroid 172974 2005 $\mathrm{YW_{55}}$} is a member of the Amor group and has delta-$v$ of 6.379 km $\mathrm{s^{-1}}$.
The geometric albedo of the asteroid is 0.30 \citep{Mueller2011a}.
Colorimetric observations for the asteroid were performed during its apparition in February 2008.
The colorimetric data obtained in this study yielded classification of this asteroid as S/D-type (Fig. \ref{fig:S1}).
The classification of this asteroid could be constrained to S-type through consideration of the albedo.

\subsubsection{190491 2000 $\mathbf{FJ_{10}}$} 
Asteroid 190491 2000 $\mathrm{FJ_{10}}$ with a delta-$v$ of 4.553 km $\mathrm{s^{-1}}$ belongs to the Apollo group.
\citet{Christou2012} reported that the asteroid has an S-type surface and a rotational period of more than 2 hr.
Lightcurve observations were performed for 2000 $\mathrm{FJ_{10}}$ during its apparition in August 2014.
A rotational period of 2.456 hr was obtained (Fig. \ref{fig:L1}), which is consistent with the literature.

\subsubsection{206378 2003 RB}
Asteroid 206378 2003 RB is a member of the Apollo group and has a delta-$v$ of 5.510 km $\mathrm{s^{-1}}$.
\citet{Warner2016a} showed that the asteroid has a rotational period of 37.5 hr, with an albedo of 0.44 \citep{Nugent2016}.
Colorimetric observations of 2003 RB were performed during its apparition in September 2009, and it was classified as S/A-type.
Moreover, the combination of the visible data acquired in this study and the near-infrared data in the SMASS database\footnotemark[8] allowed the asteroid to be classified as Sw-type.

\subsubsection{214869 2007 $\mathbf{PA_{8}}$}
Asteroid 214869 2007 $\mathrm{PA_{8}}$ belongs to the Apollo group and has a delta-$v$ of 6.797 km $\mathrm{s^{-1}}$.
The albedo of 2007 $\mathrm{PA_{8}}$ is 0.29 \citep{Fornasier2015}. 
The asteroid has an SC/OC of 0.29 and the rotational period was shown to be 102 hr \citep{Brozovic2017}.
\citet{Brozovic2017} reported that this asteroid is in non-principal axis rotation and a short-axis mode precession with average period of 102 hr and oscillation about the long axis of 493 hr. 
Spectrophotometric and spectroscopic observations of the asteroid were conducted in \citet{Hicks2012}, \citet{Lin2018}, \citet{Nedelcu2014}, \citet{Fornasier2015}, \citet{Sanchez2015}, and \citet{Perna2016}.
Through a combination of the visible data reported by \citet{Kuroda2014} and the infrared data of the SMASS database\footnotemark[8], this asteroid was classified as Sq-type (Fig. \ref{fig:S1}). 
Several spectra including those from the literature were obtained and compared. 
The spectra are roughly coincident, but there are slight differences in its shape.

\subsubsection{225312 1996 $\mathbf{XB_{27}}$}
Asteroid 225312 1996 $\mathrm{XB_{27}}$ belongs to the Amor group and has a delta-v of 4.755 km $\mathrm{s^{-1}}$.
It has been noted that this asteroid is possibly long-lived because of its low eccentricity and inclination, and because it does not cross the orbits of either the Earth or Mars (\cite{Evans1999}, \yearcite{Evans2002}).
The albedo of the asteroid is 0.48 \citep{Mueller2011a}.
1996 $\mathrm{XB_{27}}$ was classified as being D-type \citep{Carry2016}.
Considering the albedo distribution of D-type asteroids \citep{Hasegawa2017}, the value of the albedo and the classification result are contradictory.
This asteroid was classified as S/D-type using SDSS data in this study (Fig. \ref{fig:S1}).
Taking the albedo value into consideration, the preferable classification of this asteroid is S-type.
Lightcurve observations for 1996 $\mathrm{XB_{27}}$ were performed during its apparition in August 2014.
A rotational period of 1.195 hr was obtained (Fig. \ref{fig:L1}).
Thus, 1996 $\mathrm{XB_{27}}$ is a fast rotator.

\subsubsection{250697 2005 $\mathbf{QY_{151}}$}
Asteroid 250697 2005 $\mathrm{QY_{151}}$ is a member of the Apollo group and has a delta-$v$ of 7.006 km $\mathrm{s^{-1}}$.
This asteroid was observed during its apparition in November 2010 and its rotational period was determined to be more than 11 h from its lightcurve (Fig. \ref{fig:L1}).

\subsubsection{253062 2002 $\mathbf{TC_{70}}$}
Asteroid 253062 2002 $\mathrm{TC_{70}}$ is a member of the Amor group and has a delta-$v$ of 4.886 km $\mathrm{s^{-1}}$.
\citet{Kuroda2014} reported that 2002 $\mathrm{TC_{70}}$ is Sq/Q-type.
Lightcurve observations for 2002 $\mathrm{TC_{70}}$ were performed during its apparition in February 2013.
A rotational period of $\sim$17 hr was acquired (Fig. \ref{fig:L1}).

\subsubsection{341843 2008 $\mathbf{EV_{5}}$}
Asteroid 341843 2008 $\mathrm{EV_{5}}$ has a Aten-type orbit with a delta-$v$ of 5.633 km $\mathrm{s^{-1}}$.
The asteroid albedo is 0.10 \citep{Mainzer2011} and the asteroid has an SC/OC of 0.40 \citep{Busch2011}.
The rotation period is reported to be 3.725 hr \citep{Galad2009}.
Lightcurve observations of 2008 $\mathrm{EV_{5}}$ were performed during its apparition in January and March 2009.
Because the lightcurve data were not sufficient for periodic analysis, the rotational period of the asteroid could not be determined in this study.
The value of the rotational period was cited from \citet{Galad2009} (Fig. \ref{fig:L1}).
The colorimetric and spectroscopic data of the asteroid are reported in \citet{Somers2010} and \citet{Reddy2012}, respectively.
As classification was not performed in \citet{Reddy2012}, classification was conducted in this study; hence, the asteroid was classified as B-type.
As the albedo value for 2008 $\mathrm{EV_{5}}$ lies within the albedo distribution range of B-type asteroids \citep{Hasegawa2017}, this classification result is consistent with the spectral type inferred from the albedo value.

\subsubsection{357439 2004 $\mathbf{BL_{86}}$}
Asteroid 357439 2004 $\mathrm{BL_{86}}$, with a delta-$v$ of 8.458 km $\mathrm{s^{-1}}$, belongs to the Apollo group. 
A combination of the diameter obtained from the absolute magnitude and the size determined using radar \citep{Benner2015} yields a geometric albedo value of 0.25 in this asteroid.
\citet{Benner2015} showed that 2004 $\mathrm{BL_{86}}$ is a binary system composed of a primary body and satellite of 0.35 and 0.07 km in diameter, respectively.
The asteroid was classified as V-type (\cite{Birlan2015}; \cite{Franco2015}; \cite{Reddy2015}) and the rotational period was reported to be 2.620 hr \citep{Reddy2015}.
Lightcurve observations of 2004 $\mathrm{BL_{86}}$ were conducted during its apparition in January 2015, and a rotational period of 2.572 hr was obtained (Fig. \ref{fig:L1}). 
The phase angle changed by approximately 10 degrees in three days of the observation period.
While correction was made in consideration of this, the rotation period was 2.568 hr.
The corrected result is in agreement with past reports.

\subsubsection{363505 2003 $\mathbf{UC_{20}}$}
Asteroid 363505 2003 $\mathrm{UC_{20}}$ is a member of the Aten group and has a delta-$v$ of 8.715 km $\mathrm{s^{-1}}$.
This asteroid has an albedo of 0.03 \citep{Nugent2015}, with an SC/OC value of 0.21 \citep{Benner2008}.
Colorimetric observation for 2003 $\mathrm{UC_{20}}$ was performed during its apparition in November 2005.
The asteroid was classified as B/C-type based on visible data (Fig. \ref{fig:S1}).
By combining the visible data acquired in this study and the near-infrared data in the SMASS database, the asteroid was classified as Cb-type.

\subsubsection{414990 2011 $\mathbf{EM_{51}}$}
Asteroid 414990 2011 $\mathrm{EM_{51}}$ belongs to the Apollo group and has a delta-$v$ of 5.213 km $\mathrm{s^{-1}}$.
Spectroscopic observation of the asteroid was performed during its apparition in February 2014.
The spectrum in this study indicated an A/S-type asteroid (Fig. \ref{fig:S1}).

\subsubsection{416186 2002 $\mathbf{TD_{60}}$}
Asteroid 416186 2002 $\mathrm{TD_{60}}$ with a delta-$v$ of 5.373 km $\mathrm{s^{-1}}$ is a member of the Amor group.
The albedo of this asteroid has been estimated to be 0.05 according to the NEOSurvey website\footnotemark[9].
This asteroid has an SC/OC of 0.41 \citep{Benner2008}, and it has been noted that 2002 $\mathrm{TD_{60}}$ is a tumbling asteroid with two rotational periods of  2.8513 and 6.783 hr \citep{Pravec2005}.
Lightcurve observations of 2002 $\mathrm{TD_{60}}$ were performed during its apparition in November 2006.
The short-period value of the rotational period given in \citet{Pravec2005} was utilized (Fig. \ref{fig:L1}). 
The asteroid was classified as S-type in terms of the Bus taxonomy \citep{Lazzarin2005}.
As \citep{Lazzarin2005} did not classify this asteroid using the Bus-DeMeo taxonomy, it was classified as Sw-type for that taxonomy in this study.

\subsubsection{451157 2009 $\mathbf{SQ_{104}}$}
Asteroid 451157 2009 $\mathrm{SQ_{104}}$, which has a delta-$v$ of 4.879 km $\mathrm{s^{-1}}$, is a member of the Apollo group.
Spectroscopic data of 2009 $\mathrm{SQ_{104}}$ were reported in \citet{Kuroda2014}.
A combination of the visible data of \citet{Kuroda2014} and the SMASS near-infrared data\footnotemark[8] also indicated an Sq-type asteroid (Fig. \ref{fig:S1}).
2009 $\mathrm{SQ_{104}}$ was observed during its apparition in May 2005.
A rotational period of 6.974 hr was obtained (Fig. \ref{fig:L1}).

\subsubsection{471240 2011 $\mathbf{BT_{15}}$}
Asteroid 471240 2011 $\mathrm{BT_{15}}$ has an Apollo-type orbit with a delta-$v$ of 4.975 km $\mathrm{s^{-1}}$.
The rotation period is reported as 0.109138 hr \citep{Warner2014b}, which indicates that the asteroid is a fast rotator.
The asteroid was classified as A-type \citep{Kuroda2014}.
Based on a combination of the visible data of \citet{Kuroda2014} and the SMASS infrared data\footnotemark[8], this asteroid was classified as Sw-type (Fig. \ref{fig:S1}).

\subsubsection{481394 2006 $\mathbf{SF_{6}}$}
Asteroid 481394 2006 $\mathrm{SF_{6}}$ belongs to the Aten group and has a delta-$v$ of 6.802 km $\mathrm{s^{-1}}$.
The albedo is 0.21, from the NEOSurvey website\footnotemark[9].
Colorimetric observation for the asteroid was performed during its apparition in November 2007.
The spectrophotometric data obtained in this study allowed classification of this asteroid to A/S/D-type (Fig. \ref{fig:S1}).
The classification of this asteroid could be constrained to A/S type in consideration of its albedo value.

\subsubsection{2001 $\mathbf{QC_{34}}$}
Asteroid 2001 $\mathrm{QC_{34}}$ with a delta-$v$ of 4.972 km $\mathrm{s^{-1}}$ belongs to the Apollo group.
The albedo of this asteroid is estimated as 0.27 on the NEOSurvey website\footnotemark[9].
Colorimetric and spectroscopic data for the asteroid were obtained by \citet{Dandy2003} and \citet{Carry2016} and \citet{Vilas2008}, respectively.
Photometric observations for 2001 $\mathrm{QC_{34}}$  were performed during its apparition in September 2007. 
From the spectrophotometric data in this study, the asteroid was categorized as S/X-type (Fig. \ref{fig:S1}).
Using spectroscopic data \citep{Vilas2008}, this asteroid was classified as Q-type.

\subsubsection{2004 $\mathbf{DK_{1}}$}
The asteroid 2004 $\mathrm{DK_{1}}$ is a member of the Amor group and has the delta-$v$ of 5.627 km $\mathrm{s^{-1}}$.
Colorimetric observations of 2004 $\mathrm{DK_{1}}$ were performed during its apparition in April 2004.
This asteroid was classified as S/D/A-type. 

\subsubsection{2004 $\mathbf{QJ_{7}}$}
Asteroid 2004 $\mathrm{QJ_{7}}$ belongs to the Apollo group and has a delta-$v$ of 6.159 km $\mathrm{s^{-1}}$.
The asteroid was classified as Q-type \citep{DeMeo2014}.
2004 $\mathrm{QJ_{7}}$ was observed during its apparition in November 2011 and a rotational period of 1.28 hr was acquired (Fig. \ref{fig:L1}).
This asteroid is a fast rotator.

\subsubsection{2004 $\mathbf{XL_{14}}$}
Asteroid 2004 $\mathrm{XL_{14}}$ belongs to the Aten group and has a delta-$v$ of 12.436 km $\mathrm{s^{-1}}$.
The albedo is 0.15 according to the NEOSurvey website\footnotemark[9]. 
The SC/OC is 0.49 \citep{Benner2008}.
Colorimetric observations of 2004 $\mathrm{XL_{14}}$ were conducted during its apparition in December 2006.
The spectrophotometric data obtained in this study allowed classification of this asteroid as X/D/S-type (Fig. \ref{fig:S1}).
Moreover, the asteroid was classified as Xk-type based on a combination of the visible data in this study and the infrared data of the SMASS database\footnotemark[8]. 

\subsubsection{2005 $\mathbf{JU_{108}}$}
Asteroid 2005 $\mathrm{JU_{108}}$ is a member of the Amor group and has a delta-$v$ of 8.436 km $\mathrm{s^{-1}}$.
This asteroid has been classified as C-type \citep{Carry2016}.
Lightcurve observations for 2005 $\mathrm{JU_{108}}$ were performed during its apparition in August 2015, and a rotational period of 5.34 hr was obtained (Fig. \ref{fig:L1}).

\subsubsection{2005 TF}
Asteroid 2005 TF is a member of the Amor group and has a delta-$v$ of 5.899 km $\mathrm{s^{-1}}$.
A rotational period of 2.8692 hr has been reported for this asteroid \citep{Warner2017a}.
2005 TF was observed during its apparition in November and December 2005.
The rotational period could not be obtained in this study because insufficient data were acquired for periodic analysis.
Thus, the value of the rotational period was quoted from \citet{Warner2017a} (Fig. \ref{fig:L1}).
The spectrophotometric data in this study allowed classification of this asteroid as Q-type (Fig. \ref{fig:S1}).
This asteroid was also classified as Sqw-type based on a combination of the visible data acquired in this study and the SMASS near-infrared data\footnotemark[8].

\subsubsection{2006 GB}
Asteroid 2006 GB has an Aten-type orbit with a delta-$v$ of 6.341 km $\mathrm{s^{-1}}$.
Photometric observation for 2006 GB was conducted during its apparition in April 2006.
A rotational period of more than 2.5 hr was obtained (Fig. \ref{fig:L1}).
The asteroid was classified as Q/X/S/D-type using the spectrophotometric data acquired in this study (Fig. \ref{fig:S1}).

\subsubsection{2007 $\mathbf{BB_{50}}$}
Asteroid 2007 $\mathrm{BB_{50}}$ with a delta-$v$ of 8.436 km $\mathrm{s^{-1}}$ belongs to the Amor group.
Colorimetric observation of 2007 $\mathrm{BB_{50}}$ was performed during its apparition in February 2007.
This asteroid was classified as S/D-type (Fig. \ref{fig:S1}). 

\subsubsection{2007 $\mathbf{BJ_{29}}$}
Asteroid 2007 $\mathrm{BJ_{29}}$ is a member of the Apollo group and has a delta-$v$ of 9.970 km $\mathrm{s^{-1}}$.
Colorimetric observation for the asteroid was performed during its apparition in February 2007.
The asteroid was classified as Q-type based on the visible data acquired in this study (Fig. \ref{fig:S1}).

\subsubsection{2007 $\mathbf{FK_{1}}$}
Asteroid 2007 $\mathrm{FK_{1}}$ belongs to the Amor group and has a delta-$v$ of 6.570 km $\mathrm{s^{-1}}$.
2007 $\mathrm{FK_{1}}$ was observed during its apparition in May 2007.
The rotational period of the asteroid was determined as 17.11 hr from its lightcurve (Fig. \ref{fig:L1}).

\subsubsection{2007 $\mathbf{RV_{9}}$}
Asteroid 2007 $\mathrm{RV_{9}}$ with a delta-$v$ of 6.624 km $\mathrm{s^{-1}}$ is assigned to the Apollo group.
Colorimetric observations of 2007 $\mathrm{RV_{9}}$ were conducted during its apparition in February 2008.
The spectrophotometric data obtained in this study indicated an asteroid of Q/S-type (Fig. \ref{fig:S1}).

\subsubsection{2007 $\mathbf{TU_{24}}$}
Asteroid 2007 $\mathrm{TU_{24}}$, which has a delta-$v$ of 6.093 km $\mathrm{s^{-1}}$, is a member of the Apollo group.
Rotational periods of $\sim$26 hr are reported on the Pravec website\footnotemark[10].
The SC/OC of the asteroid is 0.37 \citep{Benner2008}. 
Photometric data in the V and $R_{\rm C}$ bands are reported in \citet{Betzler2008}, and 2007 $\mathrm{TU_{24}}$ was colorimetrically observed during its apparition in February 2008.
The spectrophotometric data obtained in this study and those reported in \citet{Betzler2008} allowed classification of this asteroid as Q-type (Fig. \ref{fig:S1}). 

\subsubsection{2010 $\mathbf{JV_{34}}$}
Asteroid 2010 $\mathrm{JV_{34}}$ has an Apollo-type orbit with a delta-$v$ of 6.754 km $\mathrm{s^{-1}}$.
The geometric albedo of the asteroid is 0.17 \citep{Mainzer2014}.
Photometric observations for 2010 $\mathrm{JV_{34}}$ were conducted during its apparition in May 2005.
A rotational period of 2.783 hr was acquired (Fig. \ref{fig:L1}).
The asteroid was classified as Q/C-type using spectrophotometric data in this study (Fig. \ref{fig:S1}). 
Taking the value of its albedo into consideration, this asteroid classification was preferred to be Q-type.

\subsubsection{2010 $\mathbf{TC_{55}}$}
Asteroid 2010 $\mathrm{TC_{55}}$ is a member of the Amor group and has a delta-$v$ of 8.350 km $\mathrm{s^{-1}}$.
Previously, \citet{Statler2013} reported that 2010 $\mathrm{TC_{55}}$ has a rotational period of 2.446 hr. 
Colorimetric observations for the asteroid were performed during its apparition in November 2010.
The colorimetric data acquired in this study allowed classification of this asteroid as A/S-type (Fig. \ref{fig:S1}).

\subsubsection{2013 NJ}
Asteroid 2013 NJ, having a delta-$v$ of 4.908 km $\mathrm{s^{-1}}$, belongs to the Apollo group.
A rotational period of 2.02 hr has been reported \citep{Thirouin2016}.
Spectroscopic observation for 2013 NJ was performed during its apparition in December 2013. 
Further, from the spectroscopic data obtained in this study, the asteroid was categorized as V/Q-type (Fig. \ref{fig:S1}). 
Moreover, based on a combination of the visible data obtained in this study and the near-infrared data of the SMASS database\footnotemark[8], the asteroid was classified as Q-type.

\subsection{Asteroids other than NEAs}
The asteroid results given in this section are by-products of observations of the NEAs.

\subsubsection{852 Wladilena}
Asteroid 852 Wladilena (1916 S27) is an inner main-belt asteroid.
\citet{Harris1999} showed that this asteroid has a rotational period of 4.6133 hr.
The Wladilena albedo is 0.28 \citep{Usui2011}.
Colorimetric observations of this asteroid were performed during its apparition in December 2005.
This asteroid was classified as S-type (Fig. \ref{fig:S1}).

\subsubsection{7096 Napier}
Asteroid 7096 Napier (1992 VM) is a member of the group of Mars-crossing asteroids.
Napier is reported to have a geometric albedo of 0.05 \citep{Usui2011}.
This asteroid was observed during its apparition in February 2007.
From the spectrophotometric data in this study, this asteroid was classified as C-type (Fig. \ref{fig:S1}).

\subsubsection{11739 Baton Rouge}
Asteroid 11739 Baton Rouge (1998 $\mathrm{SG_{27}}$) is a member of the Hilda dynamical group.
The albedo of this asteroid is 0.08 \citep{Grav2012}.
Near-infrared colorimetric data of the asteroid are shown in \citet{Popescu2016}.
Based on the classification in \citet{Popescu2016}, this asteroid is D/T-type (D-complex).
Lightcurve observations of Baton Rouge were conducted during its apparition in December 2007, and a rotational period of 4.8 hr was obtained (Fig. \ref{fig:L1}).

\subsubsection{19483 1998 $\mathbf{HA_{116}}$}
Asteroid 19483 1998 $\mathrm{HA_{116}}$ is an inner main-belt asteroid.
A rotational period of 2.6259 hr is reported for this asteroid on the Pravec webpage\footnotemark[10].
A rotational period of 2.619 hr was acquired (Fig. \ref{fig:L1}).
The result is consistent with those in the literature.

\subsubsection{22104 2000 $\mathbf{LN_{19}}$}
Asteroid 2000 $\mathrm{LN_{19}}$ is a middle main-belt asteroid, with an albedo of 0.13 \citep{Masiero2011}.
Near-infrared colorimetric data of the asteroid are reported in \citet{Popescu2016}.
The spectrophotometric data acquired in this study allowed classification of this asteroid as X/C-complex (Fig. \ref{fig:S1}).
Additionally, this asteroid was preferred to be X-type by considering both the albedo value and the near-infrared infrared data \citep{Popescu2016}.

\subsubsection{2006 $\mathbf{EX_{52}}$}
Asteroid 2006 $\mathrm{EX_{52}}$ has a Halley- or long-period comet-type orbit.
\citet{Licandro2018} classified the asteroid as D-type.
Colorimetric observations for 2006 $\mathrm{EX_{52}}$ were conducted during its apparition in December 2006.
The asteroid was classified as a D-type asteroid using spectrophotometric data in this study (Fig. \ref{fig:S1}).

\subsubsection{2006 $\mathbf{XQ_{56}}$}
Asteroid 2006 $\mathrm{XQ_{56}}$ belongs to the Cybele dynamical group.
The asteroid albedo has been estimated to be 0.03 \citep{Nugent2015}.
Colorimetric observations of 2006 $\mathrm{XQ_{56}}$ were performed during its apparition in February 2007.
From the colorimetric data in this study, the asteroid was categorized as D-type.

\section{Discussion}
The asteroid classification performed in this study was mainly conducted using $BVR_{\rm C}I_{\rm C}$ colorimetric data.
Using the filter system, C-type asteroids and other types can be classified in cases of low-SN data.
However, it is difficult to distinguish between S- and D-type asteroids (e.g., Bivoj: classified as S/D in this study; 1998 $\mathrm{JH_{2}}$: D in \citet{Carry2016} but S in \citet{Kuroda2014}; 2000 $\mathrm{HA_{24}}$: A/S/D in this study ; 2003 $\mathrm{SD_{220}}$: S/D and S in this study; 2004 $\mathrm{TY_{16}}$: S/D and D in this study; 1996 $\mathrm{XB_{27}}$: D in \citet{Carry2016} and S in this study).
In the Tholen and Bus taxonomies, data in the wavelength range of approximately 0.8 to 1.0 \micron\ are required in order to determine the presence or absence of S-type olivine and pyroxene characteristic absorption. 
Using data for the $I_{\rm C}$ band only, there are cases in which absorption cannot be detected.
In particular, it should be noted that there is a limit to classification in the case of low-SN data in colorimetric photometry.

It is known that there are few asteroids with periods shorter than 2.2 hr.
A possible reason for this is that rubble-pile asteroids with periods faster than 2.2 hr break apart \citep{Pravec2000}.
Therefore, fast rotators with rotational rate exceeding 2.2 hr are considered to be monoliths.
The fast rotators encountered in this study are all S-complex asteroids (1996 $\mathrm{XB_{27}}$; 2011 $\mathrm{BT_{15}}$; 2004 $\mathrm{QJ_{7}}$; 2013 NJ).
2004 $\mathrm{QJ_{7}}$ and  2013 NJ have spectra that have not been reddened, but 2011 $\mathrm{BT_{15}}$ has a very reddish spectrum even compared to Itokawa and Eros, which are expected to have ordinary chondritic materials (\cite{Trombka1997}; \cite{Nakamura2011}).
Consequently, 2011 $\mathrm{BT_{15}}$ may be made of a fragment of weathered stony-iron meteorite rather than weathered ordinary chondrite.

In this study, 1994 CC was classified as ``indeterminate''.
\citet{Reddy2011} has shown that this asteroid is an olivine-dominated NEA with known composition.
Although an A-type asteroid is olivine rich, it is characterized by strong reddening.
In contrast, 1994 CC is not weathered.
The classification of this asteroid may have been deemed indeterminate because there are no non-weathered olivine rich asteroids in the Bus-DeMeo taxonomy.
The spectrum of 1994 CC matches the spectra of brachinites, which are olivine-rich achondrites and shergottites, according to the Modeling for Asteroids (M4AST) tool\footnotemark[11] \citep{Popescu2012}.
Ultimately, 1994 CC was found to be an olivine-dominated object that had not undergone space weathering.
The surface of Itokawa is weathered, but the surface age is reported to be of the order of thousands of years (\cite{Sasaki2015}; \cite{Keller2015}; \cite{Harries2016}). 
Hence, it is presumed that the surface exposure age of 1994 CC is less than thousands of years.
%This asteroid's possession of two small satellites may support this presumption.

\footnotetext[11]{$\langle$http://m4ast.imcce.fr/index.php/index/home$\rangle$.}

2013 NJ, which is a Q-type asteroid, is characteristic in that the absorption width of 1 \micron\ extends to the shorter wavelength side.
According to the M4AST tool\footnotemark[11], Almahata Sitta is a meteorite with a spectral pattern corresponding to 2013 NJ.
Almahata Sitta mainly consists of olivine and pyroxene, which is a kind of anomalous polymict ureilite.
Also, its spectrum is similar to the spectrum of Appley Bridge, which is LL6 chondrite from the Reflectance Experiment Laboratory (RELAB) database.
Like 1994 CC, 2013 NJ has a tendency not to exhibit space weathering.
The surface exposure age of the asteroid may be estimated to be less than thousands of years.

\citet{Evans2002} showed that the candidates for long-lived asteroids in the near-Earth region are 1999YB (Sq-type), 2001 $\mathrm{SW_{169}}$ (Sw-type), and 1996 $\mathrm{XB_{27}}$ (S-type), in addition to 10302 1989 ML (E-type) \citep{Mueller2007a}, 52381 1993 HA (D-type) \citep{Perna2017b}, 138911 2001 $\mathrm{AE_{2}}$ (S-type) \citep{Thomas2014}, and 2000 $\mathrm{AE_{205}}$ (S-type) \citep{Binzel2001a}.
Previously, \citet{Rossi2009} demonstrated via calculation that the NEA rotation distribution is dispersed by the Yarkovsky-O'Keefe-Radzievskii-Paddack (YORP) effect.
The asteroid rotation periods in this region are also dispersed (2001 $\mathrm{SW_{169}}$: 69.97 h;  2001 $\mathrm{AE_{2}}$: 15.88 h \citep{Warner2018}; 1989 ML: 19 h \citep{Abe2000}; 1999 YB: 9.39 h; 1993 HA: 4.107 h \citep{Perna2017b}; 1996 $\mathrm{XB_{27}}$: 1.195 h).
\citet{Bottke2007} has noted that these asteroids may be temporarily trapped in this region.
It has been inferred that asteroids that remain longer in this area are easier to spin up and down, because they experience more YORP effects.
As they have extreme rotation periods, 1996 $\mathrm{XB_{27}}$ and 2001 $\mathrm{SW_{169}}$ are likely to be long-lived asteroids.
The YORP effect depends strongly on the pole orientation.
If they are long-lived they may be driven to the stable obliquity states which could be tested.

The spectral distributions of asteroids having both low delta-$v$ of less than 6 km $\mathrm{s^{-1}}$ and absolute magnitudes not exceeding 21, while being reachable by the Hayabusa2 spacecraft, were extracted in this work.
For this purpose, the classification catalog created in this study and those of the following works were used: \citet{Binzel2001a}; \citet{Binzel2004a}; \citet{Binzel2004b}; \citet{Carry2016}; \citet{Christou2012}; \citet{Clark2011}; \citet{Dandy2003}; \citet{deLeon2010}; \citet{DeMeo2014}; \citet{Hasegawa2017}; \citet{Hicks2012}; \citet{Ieva2018}; \citet{Kuroda2014}; \citet{Lazzarin2005}; \citet{Lin2018}; \citet{Moskovitz2013}; \citet{Perna2014}; \citet{Perna2016}; \citet{Perna2018}; \citet{Tholen1984}; \citet{Thomas2014}; \citet{Xu1995}.
The spectral classification procedure provided a total of 91 asteroids, composed of 11 C-complexes, 60 S-complexes, and 20 others (11 C-complex, 6 D-complex, and 3 V-type).
Number of the S-complex, C-complex, and other asteroid were found to be 66, 12, and 22, respectively.
The results are consistent with those of \citet{Kuroda2014}, within the statistical error range, and imply that the NEAs with low delta-$v$ reachable by Hayabusa2 originate from the specific inner main-belt region, which consists of a $\nu_{6}$ secular resonance vicinity, a 3:1 mean-motion resonance vicinity, and an intermediate-source Mars-crossing region.
In particular, the NEAs appear to have been supplied through the $\nu_{6}$ resonance.
This is supported by \citet{Bottke2002}.

Detailed analysis of the Hayabusa2-reachable asteroids classified as C- and S-complexes revealed that there are six B-types and five C-types among the C-complexes and 33 Q-types among the S-complexes, including Sq-types, 17 S-types, and 10 others.
The number of B-types is approximately one tenth that of the C-types (based on \cite{Hasegawa2017}), and the number of Q-types is less than five percent the number of S-types \citep{Lin2015} for the main-belt asteroids.
Obviously, the distributions of the Q/S ratio and B/C ratio of the main belt region differ among asteroids with low delta-$v$.
In accordance with this result, the spectrum of an asteroid in the near-Earth region turns bluish.
It is believed that the spectrum changes from a reddish (S-type) spectrum to bluish (Q-type) as the surface layer is refreshed through various mechanisms (e.g., \cite{Binzel2010}; \cite{Polishook2014}; \cite{Delbo2014}).
However, the surface-layer refreshing mechanism cannot explain the change of B/C-type, as the spectra of aqueously altered carbonaceous chondrites have been found to be bluish under weathering in experiment (\cite{Matsuoka2015}; \cite{Lantz2017})

A large range of thermal inertia values is observed within the similar size ranges between 10 and 100 km in diameter, and this result implies that asteroids with sizes exceeding 10 km are covered in fine and mature particles \citep{Hanus2018}.
In addition, NEAs not covered with fine regoliths were found.
By combining the maximum polarization degree and the geometric albedo, \citet{Ishiguro2017} and \citet{Ito2018} showed the existence of particles with diameters exceeding several hundreds of micrometers on asteroid surfaces.
\citet{Binzel2015} argued that the redness and blueness of the spectrum of asteroid 101955 Bennu indicate finer and coarser grain sizes, respectively, based on laboratory measurements.
From observation with the UH88 Telescope, Ryugu was classified as C-type; however, this asteroid was classified as B-type based on LOT and Steward observations.
It was also shown that 162173 Ryugu has B- and C-type spectra on the surface \citep{deLeon2018}.
In addition, some asteroids (Hephaistos; Itokawa; 1989 UQ; 2001 $\mathrm{FC_{7}}$; 2001 $\mathrm{WC_{47}}$; 2001 $\mathrm{SN_{263}}$; 2007 $\mathrm{PA_{8}}$) were seen to have varying spectra.
In fact, it has been confirmed that the grain size of the 25143 Itokawa surface differs \citep{Fujiwara2006}.
This indicates that these objects have surface variability.
In particular, the spectra of 2001 $\mathrm{FC_{7}}$, 2001 $\mathrm{SN_{263}}$, and 2007 $\mathrm{PA_{8}}$ are both blue and red.
Similarly to Bennu, these spectra imply that these asteroids have separate surface regions dominated by finer and coarser grain materials or boulders.
Existence of coarser grain material on the surface of NEAs may explain the B/C-type ratio of NEAs.

On the other hands, the Polana-Eulalia family of the inner asteroid belt, which has low eccentricity and phase angle, has an almost equal proportion of B- and C-type asteroids \citep{deLeon2018}.
Previously, \citet{Walsh2013} suggested that the major contributors of primitive NEAs could be the Eulalia and new Polana family.
It is also possible that most of the C-complex NEAs accessible by Hayabusa2 are from this family.

\section{Conclusions}
Observations of 74 NEAs have elucidated the physical characteristics of low delta-{\it v} bodies for sample return missions such as Hayabusa2.

\begin{description}
\item{$\bullet$}
Among the asteroids observed in this study, the C-type asteroids reachable by Hayabusa2 are 153591 2001 $\mathrm{SN_{263}}$ and 341843 2008 $\mathrm{EV_{5}}$.
\item{$\bullet$}
In this study, it was found that visible colorimetric observations such as the $BVR_{\rm C}I_{\rm C}$ filter system are not very suitable for taxonomic classification, especially for classification of D- and S-type asteroids.
\item{$\bullet$}
The B/C ratio of the near-asteroids is equal to approximately one and differs from that of the asteroids in the main belt.
There are two plausible reasons for this difference.
One is related to the particle size of the regoliths on the asteroid surfaces, and the other is that the asteroids are reflected to the composition of the family from which the asteroid originated.

\end{description}

\bigskip

\begin{ack}
We would like to thank Dr. E. Howell for her careful and constructive reviews, which helped us improve the manuscript significantly.
%We are grateful to Dr. T. Hiroi for sharing meteorite spectra.
We are grateful to Dr. T. Hiroi and Dr. K. Ohtsuka for insights and fruitful discussions.
This study was based in part on data collected at Okayama Astrophysical Observatory and using the Subaru Telescope, which is operated by the National Astronomical Observatory of Japan.
Data collected at Kiso Observatory and using the Subaru Telescope are used, as well as data obtained from SMOKA, which is operated by the Astronomy Data Center, National Astronomical Observatory of Japan. 
Some of the data utilized in this publication were obtained and made available by the MIT-UH-IRTF Joint Campaign for NEO Reconnaissance. 
The IRTF is operated by the University of Hawaii under cooperative agreement no. NCC 5-538, with the NASA, Office of Space Science, Planetary Astronomy Program. 
The MIT component of this work was supported by NASA, grant no. 09-NEOO009-0001, and by the National Science Foundation, grant nos. 0506716 and 0907766.
The taxonomic results presented in this work were determined using a Bus-DeMeo taxonomy classification web tool developed by Dr. S. M. Slivan at MIT, with the support of the National Science Foundation (grant no. 0506716) and NASA (grant NAG5-12355).
This study has utilized the Asteroid Orbital Elements Database (Astorb DB) services operated at Lowell Observatory in Flagstaff, USA; the Modeling for Asteroids (M4AST) services, operated at IMCCE, in Parism, France; the Near-Earth Asteroids Database, operated at DLR, Berlin, Germany; the SIMBAD database, operated at CDS, Strasbourg, France; and the JPL HORIZON ephemeris generator system, operated at JPL, Pasadena, USA.
We are grateful to C. Kameoka, T. Matsuura, H. Nakajima, N. Onose, R. Suda, Y. Takagi, and T. Yamamoto for their support.
We wish to thank prof. H. Takami and prof. N. Arimoto for negotiation and management of the telescope observation time allocation. 
We would like to express our gratitude to the staff members of Ishigakijima Astronomical Observatory, Kiso Observatory, Lulin Observatory, Nayoro Observatory, Okayama Astrophysical Observatory, the Subaru Telescope, Steward Observatory, and the UH88 Telescope for their assistance.  
MI was supported by the National Research Foundation of Korea (NRF), which is funded by the South Korean government (MEST) (grant no. 2015R1D1A1A01060025).
This study was supported by JSPS KAKENHI (grant nos. JP17340133, JP15K05277, JP17K05636, and JP18K03723), and by the Hypervelocity Impact Facility (former facility name: the Space Plasma Laboratory), ISAS, JAXA.
\end{ack}

%%%
% See the manual for the detail.
%%%

\end{document}